\documentclass[english]{paper}
\usepackage[letterpaper]{geometry}
\usepackage{fancyhdr}
\usepackage[T1]{fontenc}
\usepackage[utf8]{inputenc}
\usepackage{amsmath}
\usepackage{siunitx}
\usepackage{graphicx}
\usepackage{booktabs}
\usepackage{multicol}
\usepackage{float}
\usepackage{caption}
\usepackage{subcaption}
\usepackage[numbers]{natbib}
\usepackage{hyperref}
\hypersetup{pdfauthor={Wendell Smith},pdftex,colorlinks=true,
    pdfborder={0 0 0},backref=false,allcolors=blue}
\makeatletter

\date{}
\usepackage{authblk}

\makeatother

\usepackage{babel}
\begin{document}
\renewcommand\Affilfont{\itshape\small}

\author[1,2]{W. Wendell Smith}
\author[1]{Po-Yi Ho}
\author[3,1,2,4,5]{Corey S. O'Hern}

\affil[1]{Department of Physics, Yale University, New Haven, Connecticut 06520-8120, USA}
\affil[2]{Integrated Graduate Program in Physical and Engineering Biology, Yale University, New Haven, Connecticut 06520-8114, USA}
\affil[3]{Department of Mechanical Engineering and Materials Science, Yale University, New Haven, Connecticut 06520-8286, USA}
\affil[4]{Department of Applied Physics, Yale University, New Haven, Connecticut 06520-8267, USA}
\affil[5]{Program in Computational Biology and Bioinformatics, Yale 
University, New Haven, Connecticut 06520, USA}

\title{Calibrated Langevin dynamics simulations of intrinsically
 disordered proteins} 
\maketitle
\begin{abstract}
We perform extensive coarse-grained (CG) Langevin dynamics simulations
of intrinsically disordered proteins (IDPs), which possess fluctuating
conformational statistics between that for excluded volume random
walks and collapsed globules. Our CG model includes repulsive steric,
attractive hydrophobic, and electrostatic interactions between
residues and is calibrated to a large collection of single-molecule
fluorescence resonance energy transfer data on the
inter-residue separations for $36$ pairs of residues in five IDPs:
$\alpha$-, $\beta$-, and $\gamma$-synuclein, the
microtubule-associated protein $\tau$, and prothymosin $\alpha$. We
find that our CG model is able to recapitulate the average inter-residue
separations regardless of the choice of the hydrophobicity scale,
which shows that our calibrated model can robustly capture the
conformational dynamics of IDPs. We then employ our model to study
the scaling of the radius of gyration with chemical distance in $11$
known IDPs. We identify a strong correlation between the distance to the
dividing line between folded proteins and IDPs in the mean charge and
hydrophobicity space and the scaling exponent of the radius of
gyration with chemical distance along the protein.

\end{abstract}

\begin{multicols}{2}

\section{Introduction\label{sec:Introduction}}

Intrinsically disordered proteins (IDPs) do not possess well-defined
three-dimensional structures as globular proteins do. Instead, they
display highly fluctuating conformational dynamics with little or no
persistent secondary structure in physiological conditions~\citep{Vucetic2003}.
IDPs are more expanded than collapsed globules, but more compact than
self-avoiding random coils~\citep{Sugase2007}. Because IDPs are
structurally disordered and sample many different conformations, they
can interact and bind to a wide variety of targets and participate
in many important cellular processes~\citep{Dyson2005}. A number of
studies have also shown that IDPs can aggregate to form oligomers and
fibrils that are rich in $\beta$-sheet secondary structure and linked
to the development of amyloid diseases such as Parkinson's and Alzheimer's
disease~\citep{Dobson2003,Hashimoto2001}.

There has been a significant research effort aimed at experimentally
measuring and modeling the conformational dynamics of single
IDPs. Although x-ray crystallography has provided the positions of each atom
(accurate in many cases to $< 1$\AA) in thousands of folded proteins,
static representations of the atomic positions in IDPs
cannot be obtained from x-ray crystallography, and such
representations are not even meaningful for IDPs~\citep{Dunker2001}.
Alternatively, many groups have employed single-molecule fluorescence
resonance energy transfer (smFRET) to obtain the separation
distributions between specific pairs of residues for IDPs in
solution. In brief, smFRET involves exciting a donor fluorophore with
a laser, which then selectively excites an acceptor fluorophore
depending on the distance between the two labeled residues. The donor
or acceptor excitation then decays, emitting a photon. The donor and
acceptor emit two different wavelengths of light, and the ratio of the
two emitted wavelengths gives the average distance between the two residues.
To date, smFRET has been performed on tens of IDPs, but data for the
distribution of inter-residue separations has been obtained only for
several pairs of residues for each protein. In addition, small-angle
x-ray scattering (SAXS)~\citep{uversky2001, uversky2005, li2002,
 tashiro2008, rekas2010, giehm2011}, nuclear magnetic resonance
(NMR)~\citep{Dedmon2005, salmon2010}, and fluorescence correlation
spectroscopy (FCS)~\citep{Morar2001, Nath2010} have been performed on
a number of IDPs. These provide more coarse measurements of the
structure of the protein, such as the radius of gyration (or
hydrodynamic radius), which characterizes the average size of the
protein.

In a recent manuscript~\citep{Smith2012}, we introduced a physical
model to describe the fluctuating conformational dynamics of IDPs. The
motivation for the new computational model for IDPs stems in part from
the fact that commonly used molecular mechanics force fields, such as
Amber~\citep{SalomonFerrer2013} and CHARMM~\citep{Brooks2009}, can bias the simulation results toward folded
behavior since they have been calibrated using x-ray crystal
structures of folded proteins~\citep{Mittag2007}. Our physical model
includes repulsive steric interactions, screened electrostatic
interactions between charged residues, and attractive hydrophobic
interactions between $C_{\alpha}$ atoms. We employed two
representations of IDPs at different spatial scales. The united-atom (UA)
description provides a realistic atomic-level representation of
protein stereochemistry, whereas the coarse-grained (CG) description
employs one bead per residue with bond-length, bond-angle, and
backbone dihedral-angle potentials derived from interactions in the UA
description.

For both UA and CG descriptions, the model requires only one free
parameter that gives the ratio of the hydrophobic to
electrostatic energy scales. In our previous work~\citep{Smith2012},
we determined this ratio by matching Langevin dynamics simulations of
the model to experimental smFRET data for the inter-residue
separations for the IDP, $\alpha$-synuclein. We then showed that our
calibrated Langevin dynamics simulations for $\alpha$-synuclein were
able to accurately recapitulate SAXS measurements of the radius of
gyration and give conformational statistics that are intermediate
between random walk and collapsed globule behavior. An advantage of
our {\it calibrated} Langevin dynamics simulations over constraint
methods is that they do not assume random walk statistics with
artificial constraints imposed on the inter-residue separation
distributions~\citep{nath}.

In this manuscript, we present extensive new results on the CG 
description of IDPs. We improve the calibration of the CG model by 
considering a larger dataset of smFRET results from experiments that 
includes five IDPs: $\alpha$-, $\beta$-, and $\gamma$-synuclein 
($\alpha$S, $\beta$S, and $\gamma$S), the microtubule-associated 
protein $\tau$ (MAPT), and prothymosin $\alpha$ (ProT$\alpha$). For 
this set of proteins, there is smFRET data on a total of $36$ pairs of 
residues ($\alpha$S: $12$~\citep{Trexler2010,trexler2013}; $\beta$S and 
$\gamma$S: $5$ each~\citep{ducas2014}; MAPT: $12$~\cite{nath}; and 
ProT$\alpha$: $2$~\citep{Hofmann2012}), which includes most of the 
smFRET data that is currently available for IDPs. In future work, our CG 
Langevin dynamics simulations can be employed to study association, 
aggregation, and formation of $\beta$-strand order in systems 
containing multiple IDPs.

IDPs typically possess low mean hydrophobicity and high mean charge
relative to folded proteins, with a dividing line in
charge-hydrophobicity space that separates the
two~\citep{Uversky2000,Mao2010}. The synucleins and MAPT are both
close to the dividing line, whereas ProT$\alpha$ is highly
charged with relatively low hydrophobicity, and is in this sense an
ideal IDP. What physical properties distinguish IDPs that are close
versus far from the folded protein/IDP dividing line? In this study, we
perform calibrated Langevin dynamics simulations of CG descriptions of
IDPs to investigate the effects of hydrophobicity and charge on the
conformational statistics of IDPs. A significant result of our work
is that we find a strong correlation between the distance to the
folded protein/IDP dividing line and the scaling exponent of the
radius of gyration with chemical distance along the protein.

Our manuscript is organized as follows. In Sec.~\ref{sec:Methods}, we
describe Langevin dynamics simulations of the CG model for IDPs and
discuss important biological and physical aspects of the IDPs we
consider. In Sec.~\ref{sec:results}, we demonstrate that the calibrated
Langevin dynamics accurately recapitulate the available smFRET
and SAXS experimental data and that the results are robust to variations in how
we model hydrophobicity. We then describe our studies of the scaling
of the radius of gyration with chemical distance for a large sample of
known IDPs. Finally, in Sec.~\ref{sec:Conclusions}, we discuss the
implications of our results on future research of IDPs.

\section{Methods\label{sec:Methods}}

\begin{table*}
\begin{centering}
\begin{tabular}{cSSSSS}
\toprule
Amino acid type & {$\alpha$S} & {$\beta$S} & {$\gamma$S} & {MAPT} & {ProT$\alpha$}\\
\midrule
ALA & 19 & 18 & 16 & 34 & 11\\
ARG$^{\text{+}}$ & 0 & 2 & 2 & 14 & 2\\
ASN & 3 & 1 & 4 & 11 & 6\\
ASP$^{\text{-}}$$^{\text{r}}$ & 6 & 3 & 3 & 29 & 19\\
CYS & 0 & 0 & 0 & 2 & 0\\
GLN & 6 & 6 & 6 & 19 & 2\\
GLU$^{\text{-}}$ & 18 & 25 & 20 & 27 & 34\\
GLY & 18 & 13 & 10 & 49 & 9\\
HIS$^{\text{+}}$ & 1 & 1 & 0 & 12 & 0\\
ILE$^{\text{a}}$ & 2 & 2 & 2 & 15 & 1\\
LEU$^{\text{a}}$ & 4 & 7 & 1 & 21 & 1\\
LYS$^{\text{+}}$ & 15 & 11 & 15 & 44 & 8\\
MET & 4 & 4 & 2 & 6 & 1\\
PHE$^{\text{a}}$ & 2 & 3 & 2 & 3 & 0\\
PRO$^{\text{r}}$ & 5 & 8 & 2 & 43 & 1\\
SER & 4 & 6 & 10 & 45 & 4\\
THR & 10 & 7 & 10 & 35 & 6\\
TRP$^{\text{a}}$ & 0 & 0 & 0 & 0 & 0\\
TYR & 4 & 4 & 1 & 5 & 0\\
VAL & 19 & 13 & 21 & 27 & 5\\
\midrule
Total & 140 & 134 & 127 & 441 & 110\\
\bottomrule
\end{tabular}
\par\end{centering}

\protect\caption{\label{Flo:composition}Numbers of each amino acid type 
in $\alpha$S, $\beta$S, $\gamma$S, MAPT, and ProT$\alpha$. `+' and `-' 
denote positively and negatively charged residues, respectively 
(Table~\ref{Flo:charges}). `a' and `r' indicate highly hydrophobic 
($\epsilon_{i} \sim 1$) and hydrophilic ($\epsilon_{i} \sim 0$) 
residues using the scaled and shifted Monera hydrophobicity scale 
described in Sec.~\ref{sub:Hydrophobicity-models}.}

\end{table*}

\begin{figure*}
\begin{centering}
\includegraphics[width=6.5in]{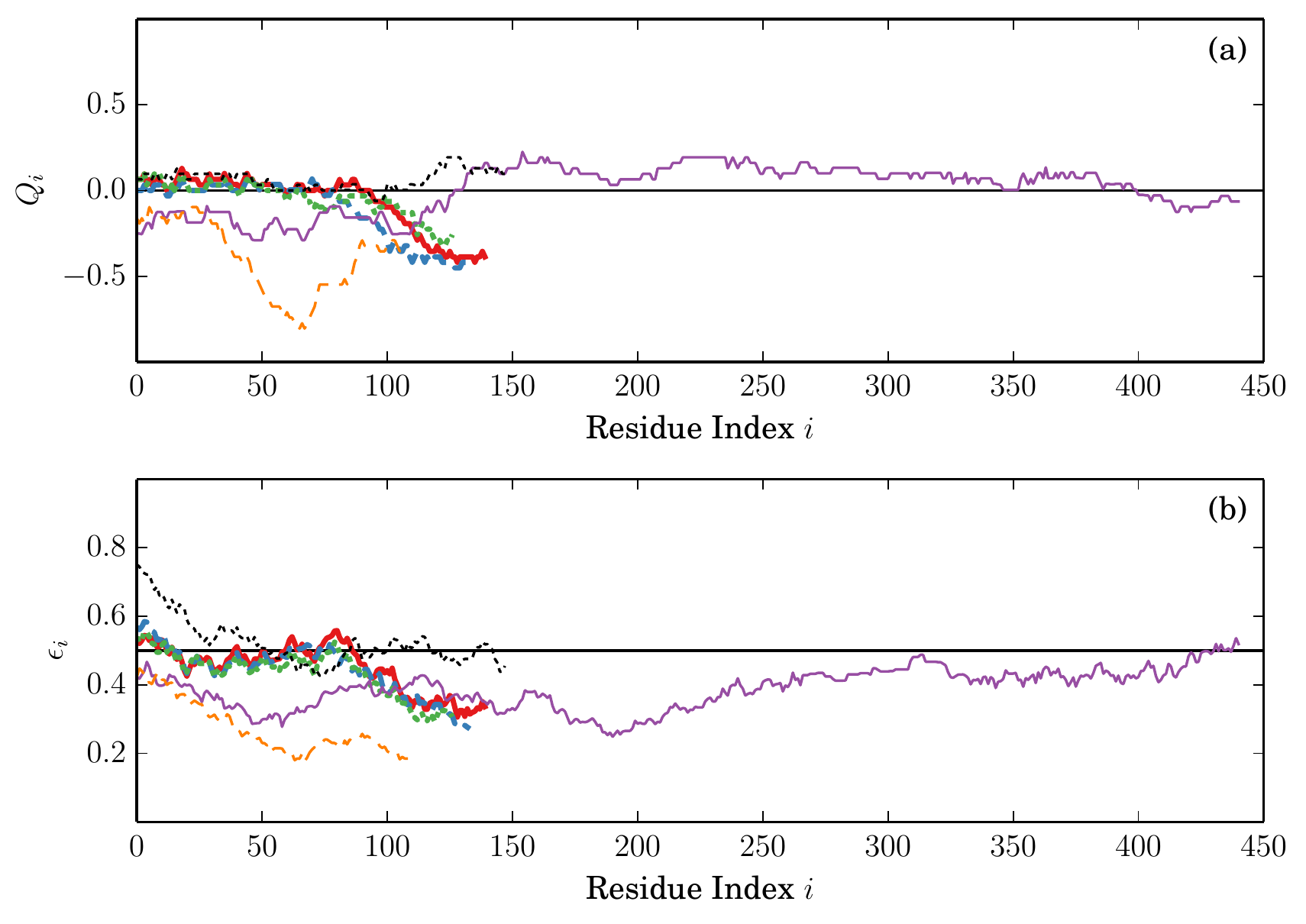}
\par\end{centering}

\protect\caption{\label{fig:chain}(Color online) (a) Electric charge 
$Q_i$ (in units of the electron charge $q_e$) and (b) hydrophobicity 
$\epsilon_i$ as a function of the residue index $i$ originating from 
the N-terminus for the IDPs $\alpha$S (thick, solid red line), $\beta$S 
(thick, dashed blue line), $\gamma$S (thick, dotted green line), MAPT 
(thin, solid purple line), ProT$\alpha$ (thin, dashed orange line) and 
the folded protein lysozyme C (thin, dotted black line). We quote the 
normalized and shifted Monera hydrophobicity scale~\citep{Monera1995}, 
where $0$ is the least and $1$ is the most hydrophobic. (See 
Eq.~\ref{normalize}.) Data for each $i$ is averaged over $31$ nearby 
residues, with data at the endpoints reflected beyond the endpoints to 
reduce edge effects.}

\end{figure*}

This manuscript focuses on the conformational dynamics of IDPs, 
including the synuclein family ($\alpha$S, $\beta$S, and $\gamma$S), 
MAPT, and ProT$\alpha$. Table~\ref{Flo:composition} provides the 
numbers of each amino acid type in these five IDPs. In 
Fig.~\ref{fig:chain}, we show the hydrophobicity and electric charge 
averaged over nearby residues as a function of the residue index 
(originating at the N-terminus) for each IDP and the folded protein 
lysozyme C~\citep{Eschenfeldt1986,Castanon1988}. 

The synucleins are a family of small proteins commonly expressed in 
neuronal tissue~\citep{Hashimoto2001}. They possess hydrophilic and 
negatively charged C-terminal 
regions~\citep{Ueda1993,Jakes1994,Ji1997}. MAPT is a 
microtubule-associated protein commonly expressed in neurons 
\citep{Goedert1988}. We study isoform F of MAPT with $441$ 
residues~\citep{Goedert1989}. The N-terminus is negatively charged, 
while the remainder is nearly neutral, and most of the protein is 
slightly hydrophilic. ProT$\alpha$, with $110$ residues, is 
both highly charged and hydrophilic~\citep{Gast1995,MullerSpath2010}. 
Note that the net hydrophobicity is larger and the net charge is much 
smaller for the folded protein lysozyme C compared to the IDPs.

\subsection{Coarse-grained model\label{sub:CG-model}}

We model IDPs using a coarse-grained
description~\citep{Clementi2008} of the backbone of a protein chain, where each
residue $i$ is represented by a spherical bead with diameter $\sigma$,
mass $M_{i}$, hydrophobicity $\epsilon_{i}$, and charge $Q_{i}$. The 
bond lengths and bond angles are constrained using linear spring 
potentials: 
\begin{equation}
V^{bl}=\frac{k_{\ell}}{2}\sum_{\left\langle ij\right\rangle }\left(r_{ij}-\ell\right)^{2}\label{eq:vbl}
\end{equation}
\begin{equation}
V^{ba}=\frac{k_{\theta}}{2}\sum_{\left\langle ijk\right\rangle }\left(\theta_{ijk}-\theta_0 \right)^{2},\label{eq:vba}
\end{equation}
where $\langle ij\rangle$ ($\langle ijk\rangle$) indicates a sum over
distinct pairs (triples) of adjacent beads, $r_{ij}$ is the separation
between the centers of beads $i$ and $j$, and $\theta_{ijk}$ is the
angle between the bonded residues $i$, $j$, and $k$. The average bond
length $\langle r_{ij} \rangle = \ell$, bond angle $\langle
\theta_{ijk} \rangle = \theta_0$, and the spring constants $k_l$ and
$k_{\theta}$ in Eqs.~\ref{eq:vbl} and~\ref{eq:vba} are obtained by
calculating the average and standard deviation of $r_{ij}$ and
$\theta_{ijk}$ from Langevin dynamics of the UA model for the five
IDPs we considered with hard-sphere atomic interactions and
stereochemical constraints obtained from the Dunbrack database of
high-resolution protein crystal structures~\citep{Wang2003}. We found
$\ell=\SI{3.9}{\angstrom}$,
$(k_{\ell}/k_bT)^{-\frac{1}{2}}=\SI{0.046}{\angstrom}$, $\theta_{0} =
\SI{2.12}{radians}$, and
$(k_{\theta}/k_bT)^{-\frac{1}{2}}=\SI{0.26}{radians}$ for simulations
of $\alpha$S, where $T$ is the temperature.
Similar results are found for the other four IDPs.

\begin{table}[H]
\begin{centering}
\begin{tabular}{SSS}
\toprule
$n$ & $A_n$ & $B_n$\\
\midrule
1 & 0.705 & -0.175\\
2 & -0.313 & -0.093\\
3 & -0.079 & 0.030\\
4 & 0.041 & 0.030\\
\bottomrule 
\end{tabular}
\par\end{centering}

\protect\caption{\label{Flo:dihedrals}The four Fourier coefficients 
of the backbone dihedral angle potential $V^{da}$ in Eq.~\ref{eq:vda}
that are employed to recapitulate the probability distribution
$P^{UA}(\phi_{ijkl})$ of backbone dihedral angles (defined by four consecutive 
C$_{\alpha}$ atoms) from UA
simulations of $\alpha$S.}
\end{table}

\begin{figure*}
\begin{centering}
\includegraphics[width=4.5in]{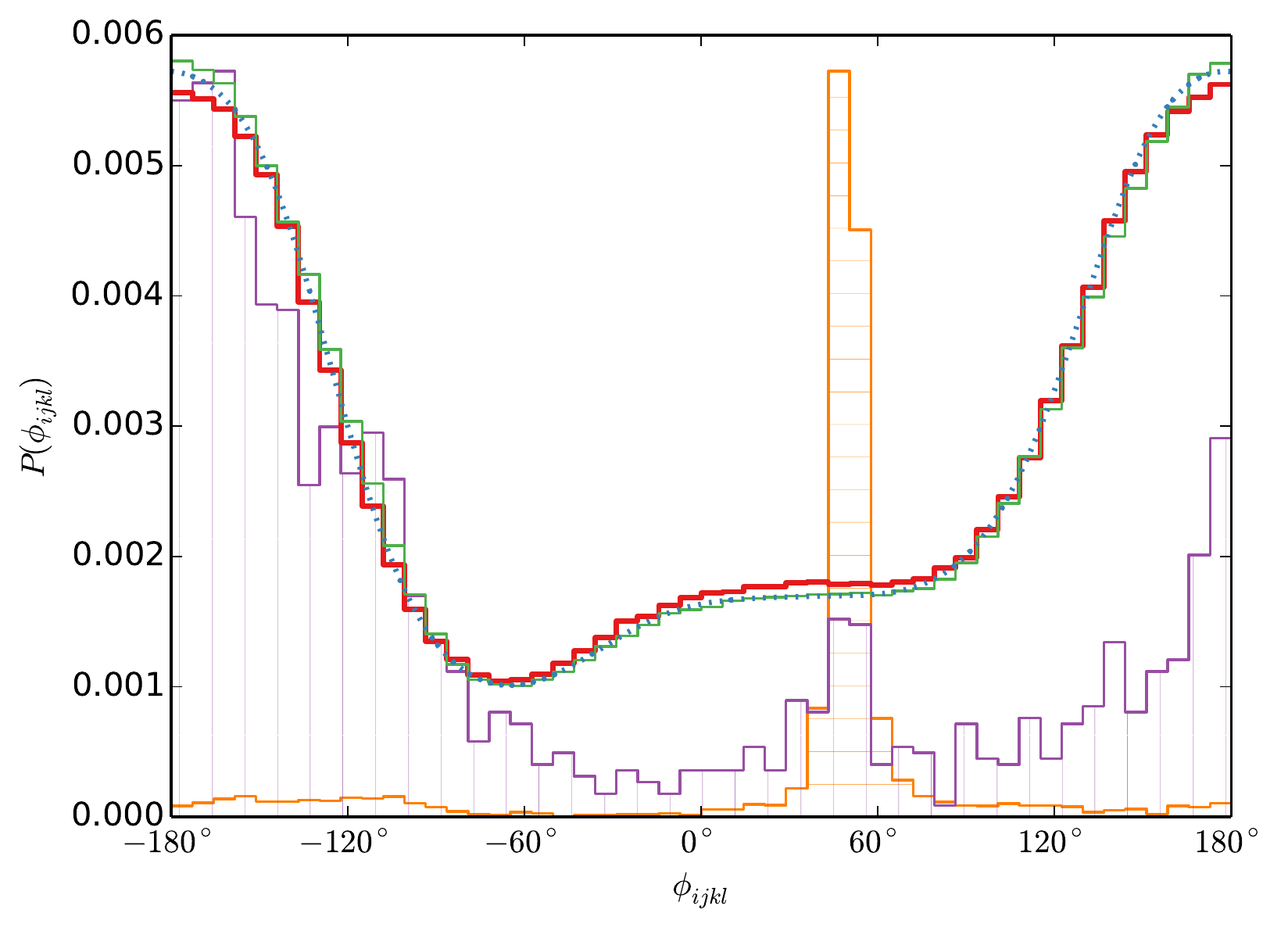}
\par\end{centering}
\protect\caption{\label{fig:dadist}(Color online) The backbone dihedral angle distribution
$P^{UA}(\phi_{ijkl})$ obtained from the UA description of
$\alpha$S (light green solid line) with only hard-sphere
atomic interactions plus stereochemical constraints obtained from
the Dunbrack database of high-resolution protein crystal
structures. We fit $P^{UA}(\phi_{ijkl})$ for the UA model using four
coefficients (Table~\ref{Flo:dihedrals}) in the Fourier series in
Eq.~\ref{eq:vda} (blue dotted line). We show that
$P^{CG}(\phi_{ijkl})$ from Langevin dynamics simulations of the CG
model for $\alpha$S with only bond-length, bond-angle, and
dihedral angle interactions in Eqs.~\ref{eq:vbl},~\ref{eq:vba},
and~\ref{eq:vda} (thick solid red line) matches that from the
hard-sphere UA model for $\alpha$-synuclein. $P(\phi_{ijkl})$ from stretches of
$\alpha$-helices (orange horizontal lines) and $\beta$-sheets
(purple vertical lines) that are longer than $10$ residues in the
Dunbrack database of high-resolution protein crystal structures are
also shown for comparison. For ease of visual comparison, the
dihedral angle distributions from $\alpha$-helical and $\beta$-sheet
structures were not normalized.
}
\end{figure*}

We show the probability distribution $P^{UA}(\phi_{ijkl})$ of backbone
dihedral angles (defined for four consecutive C$_{\alpha}$ atoms) from
Langevin dynamics simulations of the UA model for $\alpha$S
with hard-sphere atomic interactions and stereochemical constraints on
the bond-lengths and angles in Fig.~\ref{fig:dadist}. (Similar results 
are found for the other four IDPs.) The distribution possesses a large
broad peak at $\phi_{ijkl} = \pm 180^{\circ}$ and a plateau in the
range $0^{\circ} < \phi_{ijkl} < 120^{\circ}$. The broad peak and
plateau region in $P^{UA}(\phi_{ijkl})$ arise from the 
$\beta$-sheet and $\alpha$-helix backbone conformations, respectively.
We assume that a fourth-order Fourier series can describe an effective
backbone dihedral angle potential,
\begin{equation}
V^{da} = \sum_{\left\langle ijkl\right\rangle }\sum_{n=1}^{4}A_{n}\cos\left(n\phi_{ijkl}\right)+B_{n}\sin\left(n\phi_{ijkl}\right), \label{eq:vda}
\end{equation}
that governs $P^{CG}(\phi_{ijkl})$ for the CG model. In
Eq.~\ref{eq:vda}, $\langle ijkl \rangle$ indicates all distinct
combinations of four bonded residues ($i$, $j$, $k$, and $l$) along
the chain and the coefficients $A_n$ and $B_n$ are obtained by
inverting the probability distribution
$P^{UA}(\phi_{ijkl})=P^{CG}(\phi_{ijkl})$. $V^{da} = -k_b T \langle
\ln P^{CG}(\phi_{ijkl}) \rangle$ with the coefficients $A_n$ and $B_n$
given in Table~\ref{Flo:dihedrals}.

As in our previous studies~\citep{Smith2012}, we employed a purely
repulsive Weeks-Chandler-Andersen (WCA) potential, the attractive part of
the Lennard-Jones potential, and screened Coulomb potential to model
the steric, hydrophobic, and electrostatic interactions,
respectively:
\begin{multline}
\label{eq:vr}
V^{r} = \epsilon_{r}\sum_{ij}\left(4\left[\left(\frac{\sigma}{r_{ij}}\right)^{12}-\left(\frac{\sigma}{r_{ij}}\right)^{6}\right]+1\right)\\
        \Theta\left(2^{\frac{1}{6}}\sigma-r_{ij}\right)
\end{multline}
\begin{multline}
\label{eq:va}
V^{a} = \epsilon_{a}\sum_{ij}\left[\epsilon_{ij}\left(4\left[\left(\frac{\sigma}{r_{ij}}\right)^{12}-\left(\frac{\sigma}{r_{ij}}\right)^{6}\right]+1\right)\right. \\
\left. \times \Theta\left(r_{ij}-2^{\frac{1}{6}}\sigma\right)-\epsilon_{ij}\right]
\end{multline}
\begin{align}
V^{es} =& \epsilon_{es}\sum_{ij}\frac{Q_{i}Q_{j}}{q_{e}^{2}}\frac{\sigma}{r_{ij}}e^{-\frac{r_{ij}}{\lambda}},\label{eq:ves}
\end{align}
where $\Theta(x)$ is the Heaviside step function, $\sigma =
\SI{4.8}{\angstrom}$ is the average distance between the centers of
mass of neighboring residues, and $Q_i$ is the electric charge
associated with each of the charged residues LYS, ARG, HIS, ASP, and
GLU (Table~\ref{Flo:charges}). The WCA potential $V^r$ is zero for
$r_{ij} > 2^{1/6} \sigma$, the hydrophobicity potential $V^a$ includes
a $-1/r^6_{ij}$ attractive tail, and the screened Coulomb potential
$V^{es}$ is negligible beyond the screening length $\lambda$. The
mixing rule $\epsilon_{ij}$ for the (shifted and normalized)
hydrophobicity index $0 \le \epsilon_i \le 1$ for each residue $i$
will be discussed in Sec.~\ref{sub:Hydrophobicity-models}. ${\overline 
\epsilon_{es}} = \epsilon_{es}/k_bT$ is a parameter that controls the 
strength of the electrostatic interactions. Typical
experimental solution conditions with $150$ mM salt concentration,
$p\text{H}=7.4$, and temperature $T=293$~K yield
$\lambda=\SI{9}{\angstrom}$ and ${\overline
 \epsilon}_{es}=\kappa_{es}=q_{e}^{2}/(4\pi \epsilon_{0}
D\sigma k_{b}T) \approx1.485$, where $D=80$ is the permittivity of
water. For most of the simulations, we set the energy scale for the repulsive
interactions $\epsilon_{r}/k_{b}T=1$ and calibrate the ratio of
strength of the hydrophobic interactions to that of the electrostatic
interactions $\alpha_{CG} = {\overline \epsilon}_a/\kappa_{es}$
to match the smFRET data.

\begin{table}[H]
\begin{centering}
\begin{tabular}{cS}
\toprule
{Residue} & {Residue charge $Q_{i}$}\\
\midrule
LYS & 1\\
ARG & 1\\
HIS & 0.1\\
ASP & -1\\
GLU & -1\\
\bottomrule 
\end{tabular}
\par\end{centering}

\protect\caption{\label{Flo:charges}Electric charge $Q_{i}$ (in units 
of the electron charge $q_e$) for the charged
residues LYS, ARG, HIS, ASP, and GLU~\citep{Oostenbrink2004}.}
\end{table}

\subsection{Hydrophobicity models}
\label{sub:Hydrophobicity-models}

We model hydrophobic interactions between residues using the
attractive part of the Lennard-Jones potential (Eq.~\ref{eq:va}). In
this section, we describe the possible choices for assigning 
the hydrophobicity index $\epsilon_i$ to each residue and mixing rule 
$\epsilon_{ij}$ for pairwise hydrophobic interactions between
residues $i$ and $j$. 

\begin{figure*}
\begin{centering}
\includegraphics[width=4.5in]{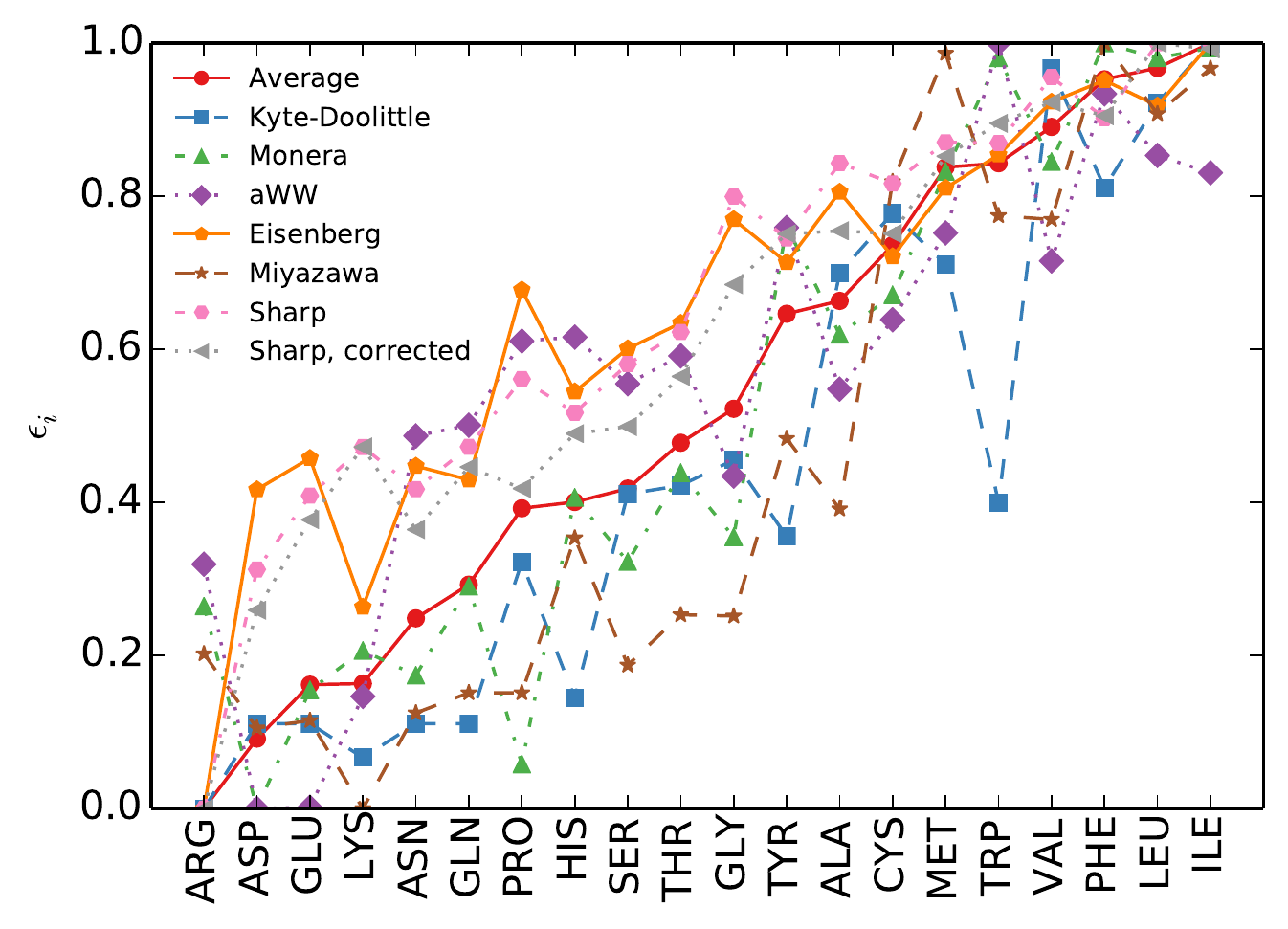}
\par\end{centering}

\protect\caption{\label{fig:hydro}(Color online) Seven commonly used 
hydrophobicity scales (Kyte-Doolittle~\citep{Kyte1982}, 
Monera~\citep{Monera1995}, Wimley-White~\citep{White1999,Wimley1996}, 
Eisenberg~\citep{Eisenberg1984}, Miyazawa~\citep{Miyazawa1985}, Sharp, 
and Sharp with solvent-solute size difference 
corrections~\citep{Sharp1991}) for each amino acid type that have been 
shifted and normalized so that $0 \le \epsilon_i \le 1$. The 
``average'' value for each residue indicates the shifted and normalized 
average over the seven shifted and normalized hydrophobicity scales. 
The residues are ordered according to their average $\epsilon_i$.} 
\end{figure*}

There are many different hydrophobicity scales for assigning the
hydrophobicity for whole residues~\citep{Cornette1987}, and each scale has its own mean,
maximum, and minimum. To enable comparison between different hydrophobicity
scales, we shifted and normalized the original values ${\widetilde
 \epsilon}_i$ to obtain
\begin{equation}
\label{normalize}
\epsilon_{i}= \frac{{\widetilde \epsilon}_i - \min_i ( {\widetilde \epsilon}_i)} { \max_i ({\widetilde \epsilon}_{i}) - \min_i ({\widetilde \epsilon}_i) } 
\end{equation}
with $0 \leq \epsilon_{i} \leq1$ as shown in Fig.~\ref{fig:hydro}. 

Below, we investigate the sensitivity of the simulation results to three
choices for the hydrophobicity scales: 1) the shifted and normalized 
Kyte-Doolittle~\citep{Kyte1982} scale, 2) the shifted and normalized Monera~\citep{Monera1995} scale, and 3)
an average of seven commonly used hydrophobicity scales
(Kyte-Doolittle, Monera, augmented
Wimley-White~\citep{White1999,Wimley1996},
Eisenberg~\citep{Eisenberg1984}, Miyazawa~\citep{Miyazawa1985}, Sharp
~\citep{Sharp1991}, and Sharp (corrected for solvent-solute size
differences~\citep{Sharp1991})). The ``average'' scale is obtained 
by averaging the seven shifted and normalized scales, and then shifting and 
normalizing the result.  

We also consider the sensitivity of the simulation results to three 
pairwise mixing rules for the shifted and normalized hydrophobicities 
$\epsilon_i$ and $\epsilon_j$: 1) arithmetic mean: 
$\epsilon_{ij}=(\epsilon_{i}+\epsilon_{j})/2$, 2) geometric mean: 
$\epsilon_{ij} = \sqrt{\epsilon_i \epsilon_j}$, and 3) maximum: 
$\epsilon_{ij} = \max(\epsilon_i,\epsilon_j)$. Below, we will show 
results (Sec.~\ref{sec:results}) for Langevin dynamics simulations of 
the five IDPs ($\alpha$S, $\beta$S, and $\gamma$S, MAPT, and 
ProT$\alpha$) using nine different models for the pairwise hydrophobic 
interactions between residues ($3$ hydrophobicity scales, each with $3$ 
mixing rules).

\subsection{Langevin dynamics simulations\label{sub:Simulation-details}}

We performed coarse-grained Langevin dynamics simulations of single
IDPs at fixed temperature $T=\SI{293}{K}$ with bond-length,
bond-angle, dihedral-angle, steric, hydrophobic, and screened Coulomb
interactions (Eqs.~\ref{eq:vbl}--\ref{eq:ves}). We employed free
boundary conditions, a modified velocity-Verlet integration scheme
with a Langevin thermostat~\citep{Allen1989}, damping coefficient
$\gamma = 0.001 \sqrt\frac{k_B T}{m_0 \ell_0^2} \approx \SI{16}{\per
 \nano \second}$, and fixed time step $\Delta t = 0.03\sqrt\frac{m_0
 \ell_0^2}{k_B T} \approx \SI{1.9}{\femto \second}$, where $m_0 = 1
{\rm Da}$ and $\ell_0 = \SI{1}{\angstrom}$. We chose the time step
$\Delta t$ so that the relative energy fluctuations in the absence of
the thermostat satisfy $\frac{\sqrt{\left<E^2\right> -
  \left<E\right>^2}}{\left<E\right>} < \num{1e-4}$ and the damping
parameter so that $1/\gamma$ is much smaller than total run time
$t_{\rm tot} =\SI{0.5}{\micro\second}$. The chains were initialized in a zig-zag
conformation with random velocities at temperature $T$ and then
equilibrated for $10^4 t_R$, where $t_R$ is the time for the
normalized $R_g$ autocorrelation function to decay to $1/e$. After
equilibration, production runs were conducted to measure the
inter-residue separations and radius of gyration for each IDP.

\section{Results\label{sec:results}}

\subsection{smFRET efficiencies}

\begin{figure*}
\begin{centering}
\includegraphics[width=6in]{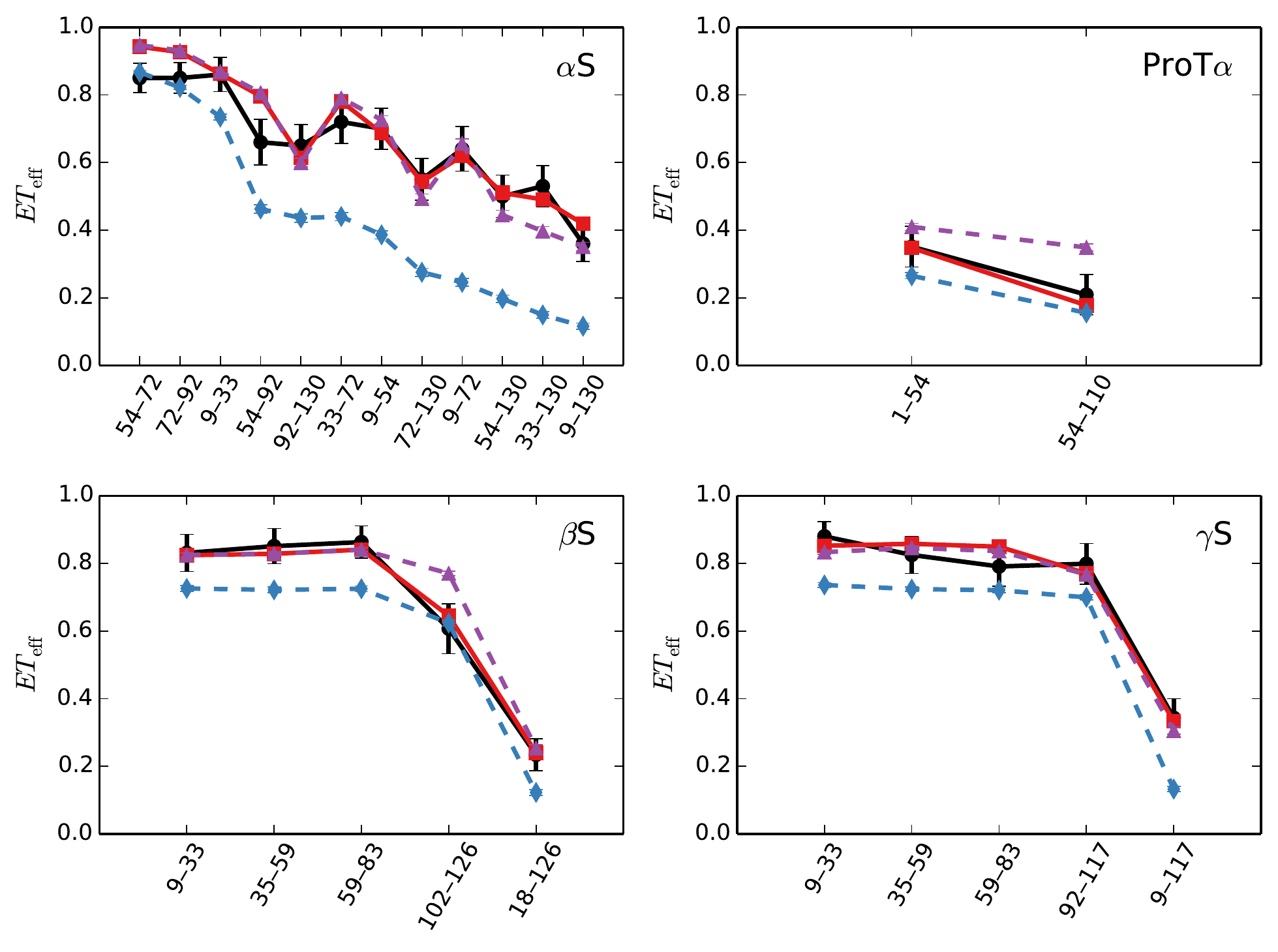}
\par\end{centering}

\protect\caption{\label{fig:smFRETaS2}(Color online) FRET efficiencies 
$ET_{\rm eff}$ for $\alpha$S (upper left), ProT$\alpha$ (upper right), 
$\beta$S (lower left), and $\gamma$S (lower right) from experimental 
measurements (black circles) and CG Langevin dynamics simulations. We 
include data for three choices for the strength of the hydrophobic and 
electrostatic interactions ${\overline \epsilon}_a$ and ${\overline 
\epsilon}_{es}$ for each IDP: 1) ${\overline \epsilon}_a=0$ and 
${\overline \epsilon}_{es} = \kappa_{es}$ such that the chains behave 
as extended coils (blue diamonds), 2) the optimal $\alpha_{CG}$ for 
each protein with ${\overline \epsilon}_{es} = \kappa_{es}$, where the 
root-mean-square deviations between the experimental and simulation 
$ET_{\rm eff}$ are minimized (red squares), and 3) the optimal 
$\alpha_{CG}$ for each protein with no electrostatic interactions 
${\overline \epsilon}_{es}=0$ (purple triangles). The error bars for 
$ET_{\rm eff}$ from the simulations give the error in the mean. Error 
bars that are not visible are smaller than the symbols.} \end{figure*}

In Fig.~\ref{fig:smFRETaS2}, we show experimental results for the
smFRET efficiencies for $12$ inter-residue separations for
$\alpha$S~\citep{Trexler2010,trexler2013}, $5$ for $\beta$S
and $\gamma$S~\citep{ducas2014}, $2$ for ProT$\alpha$~\citep{Hofmann2012},
and $12$ for MAPT~\cite{nath}. Large
$ET_{\rm eff}$ indicate small average inter-residue separations and vice
versa. The smFRET efficiencies depend strongly on the separation
$r_{ij}$ between residue pairs,
\begin{equation}
\label{ef}
ET_{\text{eff}}=\left\langle \frac{1}{1+\left(\frac{r_{ij}}{R_{0}}\right)^{6}}\right\rangle 
\end{equation}
where $R_{0}=\SI{54}{\angstrom}$ is the F\"{o}rster distance for the
donor-acceptor pair (Alexa Fluor 488--Alexa Fluor 594), angle brackets
indicate a time average, and we assume that the finite size of the
fluorophores has a negligible effect on $ET_{\rm eff}$. We directly
measure $ET_{\rm eff}$ for each IDP in our simulations, and compare it to
the experimental values. For most simulations, we varied the ratio of
the hydrophobic to the electrostatic interactions $\alpha_{CG}$ ({\it
 i.e.} change ${\overline \epsilon}_a$ at fixed ${\overline
 \epsilon}_{es} = \kappa_{es}$) to minimize the root-mean-square
deviation in $ET_{\rm eff}$ between
the simulations and experiments:
\begin{equation}
\label{rmseq}
\Delta = \sqrt{\frac{1}{N_p} \sum_{i=1}^{N_p} (ET^{\rm exp}_{\rm eff}(i) - ET^{\rm sim}_{\rm eff}(i))^2 },
\end{equation}
where $ET^{\rm exp}_{\rm eff}(i)$ is the FRET efficiency for residue pair
$i$ from experiments and $N_p$ is the number of residue pairs. For
the simulations described in this section, we employ the shifted and
normalized Monera hydrophobicity scale with the geometric mean as the
mixing rule.

\begin{figure*}
\begin{centering}
\includegraphics[width=4in]{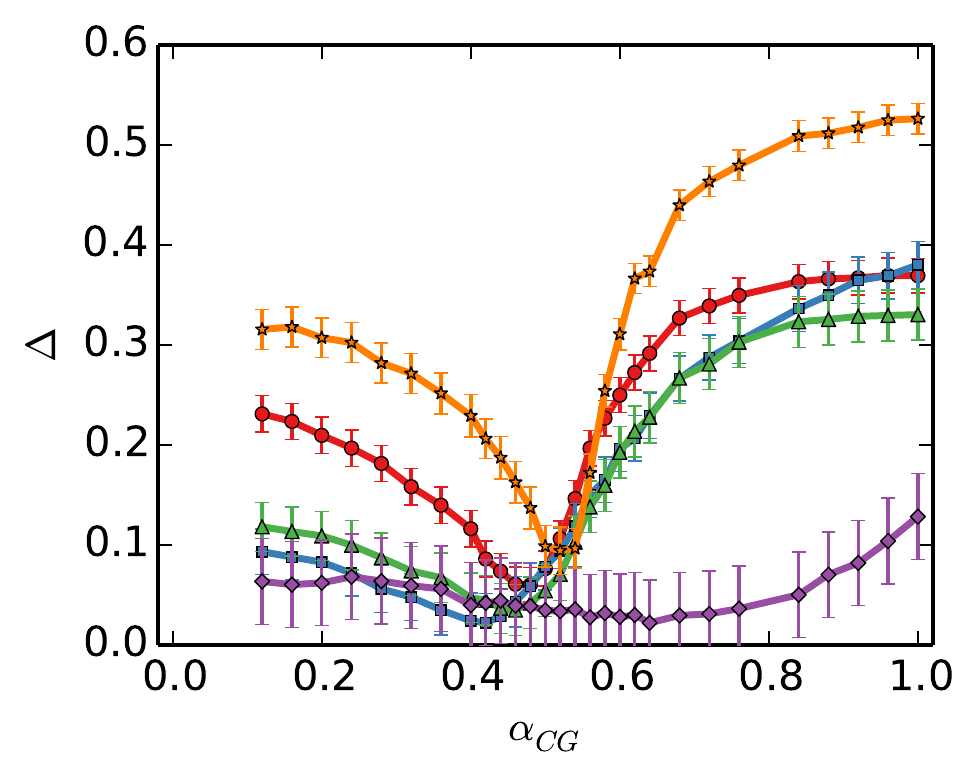}
\par\end{centering}

\protect\caption{\label{fig:delta}(Color online) Root-mean-square 
deviation $\Delta$ in $ET_{\rm eff}$ between experiments and 
simulations versus the ratio $\alpha_{CG}$ of the hydrophobicity and 
electrostatic interactions for $\alpha$S (red circles), $\beta$S (blue 
squares), and $\gamma$S (green triangles), MAPT (orange triangles), and 
ProT$\alpha$ (purple diamonds).} \end{figure*}

For $\alpha$S, the experimental values for $ET_{\rm eff}$ for the $12$ 
residue pairs vary from $\approx 0.90$ to $0.40$ as shown in 
Fig.~\ref{fig:smFRETaS2}. From the CG Langevin dynamics 
simulations of $\alpha$S, we find that $\alpha_{CG} \approx \num{0.48 
+- 0.03}$ minimizes the root-mean-square deviation between the 
simulations and experiments. This optimized value of $\alpha_{CG}$ 
yields $\Delta_{\rm min} \approx \num{0.06(2)}$ (Fig.~\ref{fig:delta}), 
which indicates close agreement between simulations and experiments. 
(The error bar for $\alpha_{CG}$ was obtained by determining the change 
in $\alpha_{CG}$ necessary for $\Delta$ to increase beyond the 
error bars of $\Delta_{\rm min}$.) In contrast, when the strength 
of the hydrophobic interactions is set to zero ($\alpha_{CG} = 0$), 
$ET_{\rm eff}$ for the simulations are significantly below the 
experimental values for all residue pairs with $\Delta \approx 
\num{0.26}$. To investigate the relative contributions of the 
hydrophobic and electrostatic interactions to $ET_{\rm eff}$, we also 
performed simulations with ${\overline \epsilon}_{es} = 0$ and 
$\alpha_{CG} = 0.48$. We find that the quality of the match between 
simulations and experiments is comparable for the simulations with 
($\Delta \approx \num{0.06(2)}$) and without ($\Delta \approx 
\num{0.08(2)}$) electrostatic interactions.

We find that $\alpha_{CG} \approx \num{0.48(3)}$ yields the best match of the
FRET efficiencies from the CG simulations and experiments for
$\alpha$S. This value of the ratio of the hydrophobic and
electrostatic interactions differs by about a factor of 2.5 from the
optimal value ($\alpha_{UA} \approx 1.2$) obtained from our previous
UA simulations of $\alpha$S~\citep{Smith2012}. This result
shows that the optimal numerical value of $\alpha$ can be sensitive to
the geometrical representation of residues in IDPs, as well as the
hydrophobicity scale implemented in the model.
 
For $\beta$S and $\gamma$S, the FRET efficiencies for three
of the five residue pairs (with similar chemical distances) are
approximately equal ($ET_{\rm eff} \approx 0.85$), while $ET_{\rm
 eff}$ for the other two pairs drop to $0.6$ and $0.2$
(Fig.~\ref{fig:smFRETaS2}). Similar to the results for $\alpha$S, the 
root-mean-square deviation in $ET_{\rm eff}$ between simulations and 
experiments is minimized when $\alpha_{CG} \approx \num{0.42 +- 0.07}$ 
and $\num{0.46 +- 0.05}$ for $\beta$S ($\Delta_{\rm min} \approx 
0.02$) and $\gamma$S ($\Delta_{\rm min} \approx 0.04$) respectively 
(Fig.~\ref{fig:delta}). In addition, the CG simulations with 
${\overline \epsilon}_{es} =0$ and $\alpha_{CG} = 0.42$ ($0.46$) show 
reasonable agreement with the experimental $ET_{\rm eff}$ for $\beta$S 
($\gamma$S) with $\Delta \approx 0.10$ ($0.04$), except for residue 
pair 102--126 for $\beta$S. For ProT$\alpha$ (Fig.~\ref{fig:smFRETaS2}), we 
find that $\Delta$ is minimized for $\alpha_{CG} \approx 
\num{0.64(36)}$. The optimal $\alpha_{CG}$ for ProT$\alpha$ has larger 
error bars because $\Delta(\alpha_{CG})$ is nearly flat. These error 
bars encompass the optimal $\alpha_{CG}$ for $\alpha$S, $\beta$S, and 
$\gamma$S.

\begin{figure*}
\begin{centering}
\includegraphics[width=4in]{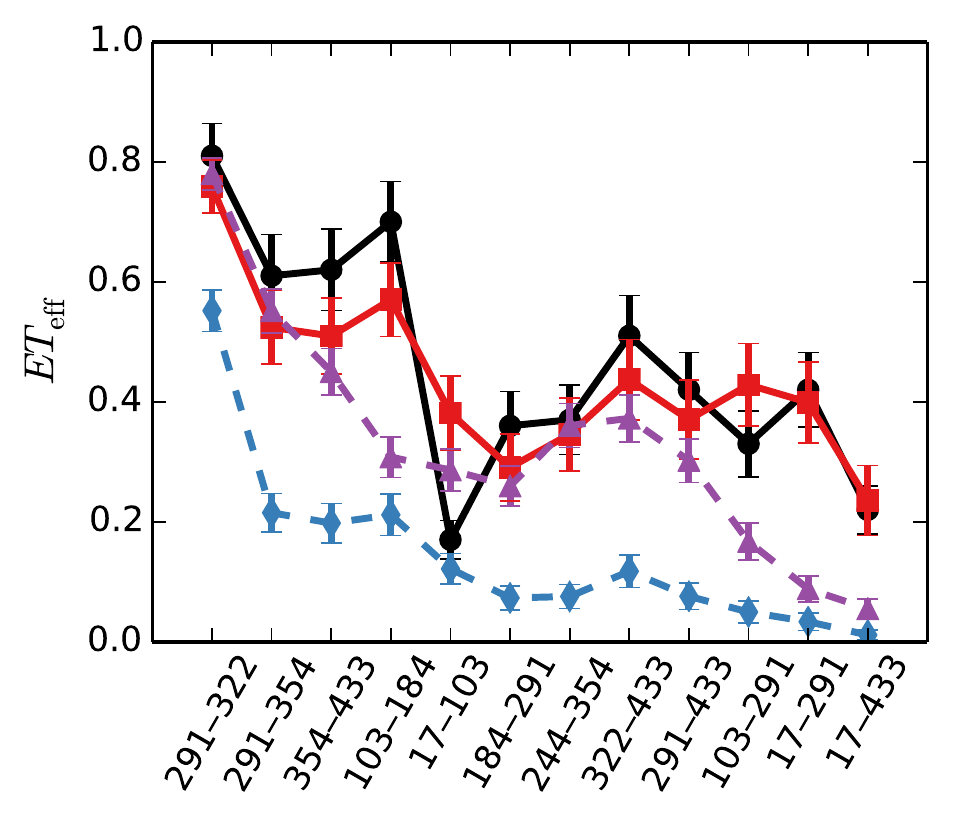}
\par\end{centering}

\protect\caption{\label{fig:tausmFRET}FRET efficiencies $ET_{\rm eff}$ 
for MAPT from smFRET experiments (black solid line with circles) and 
three CG simulations: 1) ${\overline \epsilon}_a=0$ and ${\overline 
\epsilon}_{es} = \kappa_{es}$ (blue diamonds), 2) the optimal 
$\alpha_{CG} = 0.52$ with ${\overline \epsilon}_{es} = \kappa_{es}$, 
where the root-mean-square deviations between the experimental and 
simulation $ET_{\rm eff}$ are minimized (red squares), and 3) the 
optimal $\alpha_{CG} = 0.52$ with no electrostatic interactions 
${\overline \epsilon}_{es}=0$ (purple triangles).} \end{figure*}

The $ET_{\rm eff}$ for MAPT is more complex. In 
Fig.~\ref{fig:tausmFRET}, we show $ET_{\rm eff}$ for MAPT for residue 
pairs ordered from small to large chemical distances along the protein 
chain. Despite the monotonic increase in chemical distance from left 
to right, $ET_{\rm eff}$ shows an anomalously large drop for residue 
pair 103--183 followed by an increase in $ET_{\rm eff}$ even though the 
chemical distance continues to increase. This behavior differs from the 
dependence of $ET_{\rm eff}$ on chemical distance for the synuclein 
family and ProT$\alpha$, where $ET_{\rm eff}$ decreases roughly 
monotonically with chemical distance with only minor fluctuations. In 
Fig.~\ref{fig:delta}, we show that $\Delta$ is minimized at 
$\alpha_{CG} \approx \num{0.52(2)}$. This yields $\Delta_{\min} \approx 
\num{0.09(2)}$, which is significantly larger than that for the other IDPs
(0.02, 0.02, 0.04, and 0.06).
In particular, the CG model with the optimal 
$\alpha_{CG}$ shows large deviations with the experimental $ET_{\rm 
eff}$ for residue pairs 354--433, 103--184, and 17--103. Although the 
CG model without electrostatic interactions (${\overline \epsilon}_{es} 
= 0$ and $\alpha_{CG} = 0.52$) is able to recapitulate $ET_{\rm eff}$ for the synuclein family 
and ProT$\alpha$, it yields $\Delta_{\rm min} \approx 0.18$ for MAPT. 
In the electrostatics-only CG model, we find that $\Delta_{\rm min} 
\approx 0.34$, which is much larger than $\Delta_{\rm min}$ for the CG 
model with both hydrophobic and electrostatic interactions.

\begin{figure*}
\begin{centering}
\includegraphics[width=4.5in]{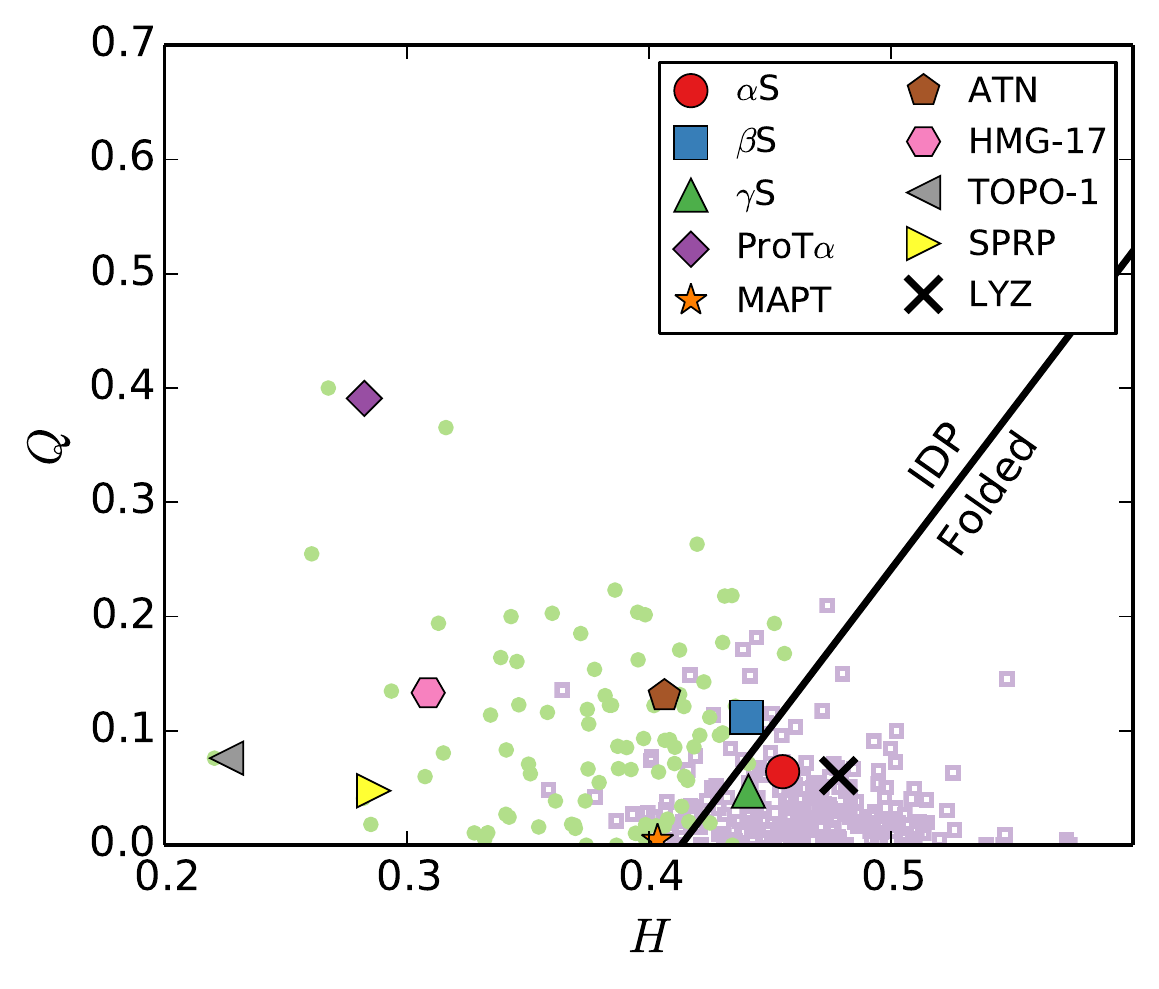}
\par\end{centering}

\protect\caption{\label{fig:qvh}Absolute value of the electric charge 
per residue $Q$ versus the hydrophobicity per residue $H$ (using the 
shifted and normalized Monera hydrophobicity scale) for known IDPs 
(small circles) and $221$ folded proteins~\citep{Uversky2000} (small 
open squares). The IDPs $\alpha$S (large circle), $\beta$S (large 
square), $\gamma$S (upward triangle), ProT$\alpha$ (diamond), MAPT 
(star), ATN (pentagon), HMG-17 (hexagon), TOPO-1 (leftward triangle), 
SPRP (rightward triangle), and the folded protein lysozyme C (X) are 
highlighted. The line $Q=2.785 H -1.151$ represents the dividing line
between IDPs (above the line) and natively folded proteins (below the 
line) given in Ref.~\citep{Uversky2000}.} \end{figure*}

We find that for the synucleins and ProT$\alpha$, optimized models with 
and without electrostatics interactions provide an accurate description 
of the experimental $ET_{\rm eff}$. However, the optimized model with 
both electrostatic and hydrophobic interactions provides the best match 
to experimental $ET_{\rm eff}$ for MAPT. More importantly, the minimal 
RMS deviations in $ET_{\rm eff}$ between simulations and experiment for 
MAPT are larger than those for the synuclein family and ProT$\alpha$ as 
well as typical experimental error bars. The fact that MAPT is three 
times as long as, and less charged and hydrophobic (Fig.~\ref{fig:qvh}) 
than the other IDPs may contribute to the larger RMS 
deviations~\citep{Szilagyi2008}.

\subsection{Sensitivity analysis of hydrophobicity models}
\label{sensitivity}

In this section, we describe results from CG Langevin dynamics
simulations of each IDP using $9$ different hydrophobicity models
(Sec.~\ref{sub:Hydrophobicity-models}): three hydrophobicity scales
(the shifted and normalized Kyte-Doolittle~\citep{Kyte1982} and 
Monera~\citep{Monera1995} scales, and an
average of seven commonly used hydrophobicity scales) plus three
pairwise mixing rules for the hydrophobicities of the residues
(arithmetic mean, geometric mean, and maximum). For the CG simulations
of each IDP and hydrophobicity model, we varied $\alpha_{CG}$ to
minimize $\Delta$.

\begin{figure*}
\begin{centering}
\includegraphics[width=4.5in]{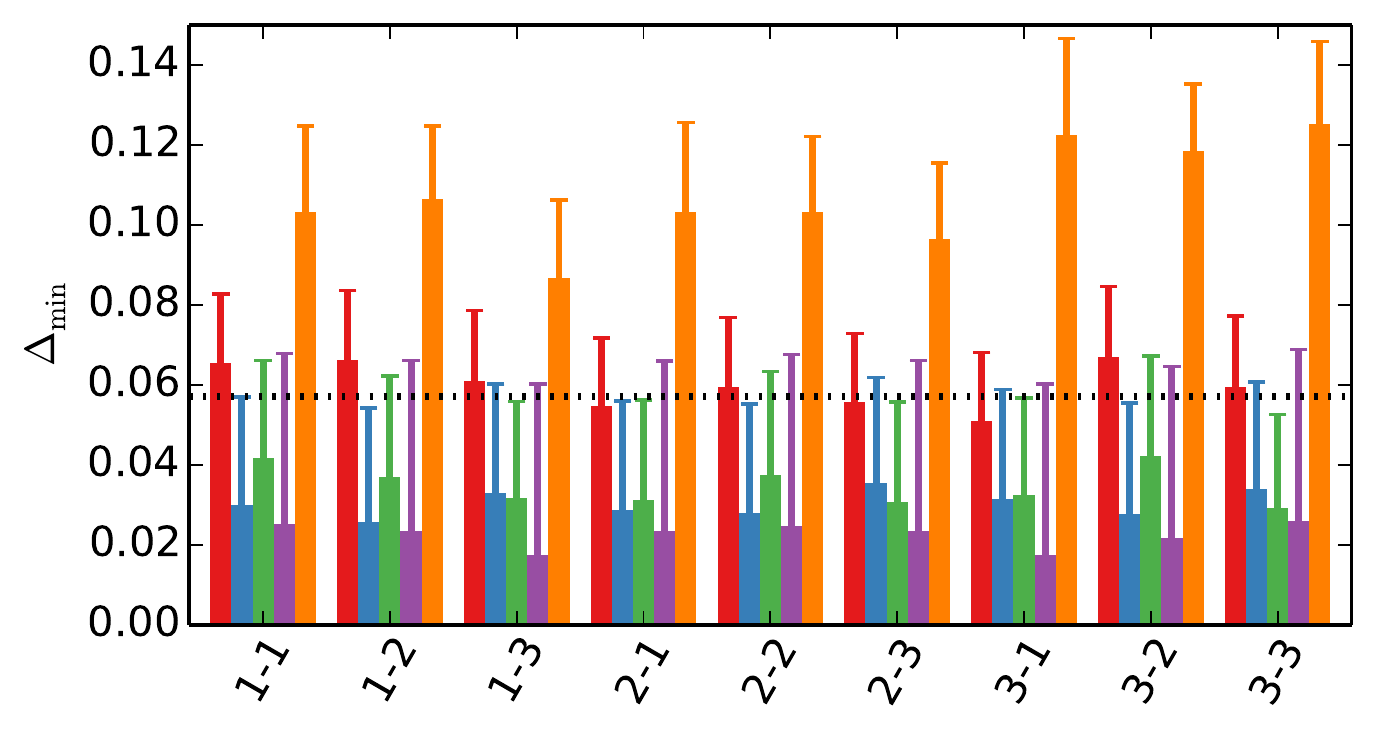}
\par\end{centering}

\protect\caption{ \label{fig:hydro2} (Color online) RMS deviations 
$\Delta_{\rm min}$in $ET_{\rm eff}$ (and its error) between the 
experiments and simulations for (from left to right) $\alpha$S (red), 
$\beta$S (blue), $\gamma$S (green), ProT$\alpha$ (purple), and MAPT 
(orange) for 9 hydrophobicity models. The labeling scheme for the nine 
hydrophobicity models is (hydrophobicity scale)-(mixing rule) as 
numbered in Sec.~\ref{sub:Hydrophobicity-models}. The RMS deviations 
are calculated from CG simulations with $\alpha_{CG}$ chosen such that 
the RMS deviations are minimized for each hydrophobicity scale and 
mixing rule. The black dotted line indicates the average error in 
$ET_{\rm eff}$ from smFRET experiments.} \end{figure*}

In Fig.~\ref{fig:hydro2}, we show the root-mean-square (RMS) deviation 
in $ET_{\rm eff}$ between the simulations and experiment and the error 
in the RMS for each IDP and hydrophobicity model. We also show an 
estimate of the average error ($6$\%) expected from the smFRET 
experiments~\citep{Trexler2010}. For most of the hydrophobicity models, 
the RMS deviations in $ET_{\rm eff}$ between simulations and experiment 
for $\beta$S, $\gamma$S, and ProT$\alpha$ are below the experimental 
error. For $\alpha$S, most of the hydrophobicity models possess RMS 
deviations that are comparable to the experimental error. Thus, for the 
synuclein family and ProT$\alpha$, the RMS deviations are comparable or 
below experimental error and the hydrophobicity model does not strongly 
affect the results. For MAPT, the RMS deviations vary from $\Delta_{\rm 
min} \approx 0.08$ to $0.12$ indicating that some of the hydrophobicity 
models are slightly better than others for this IDP.

\begin{figure*}
\begin{centering}
\includegraphics[width=4.5in]{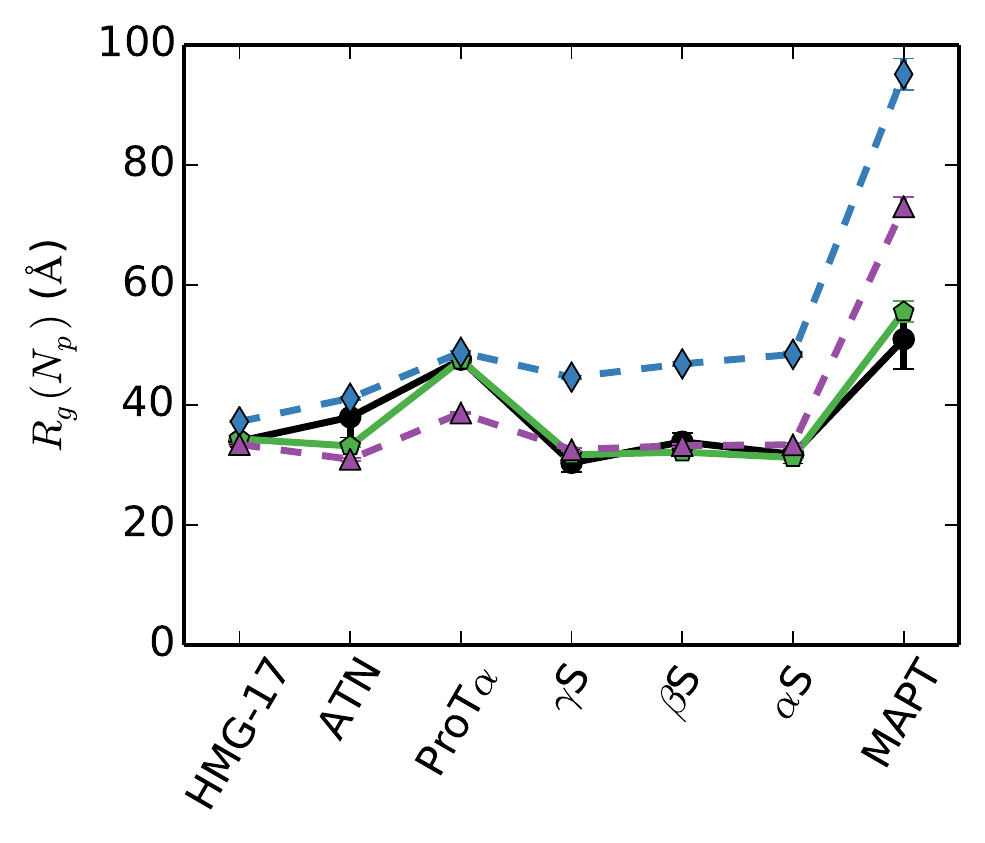}
\par\end{centering}

\protect\caption{\label{fig:rgs}Radius of gyration $R_{g}(N_{p})$ of seven 
IDPs from experiments~\citep{Johansen2011, Abercrombie1978, Gast1995,
Uversky2002b, nath} (black circles) and simulations of three CG models: 
1) $\epsilon_a = 0$ and ${\overline \epsilon}_{es} = \kappa_{es}$ such 
that the chains behave as extended coils (blue diamonds), 2) 
$\alpha_{CG}=\num{0.50}$ and ${\overline \epsilon}_{es} = 
\kappa_{es}$ (green pentagons), and 3) $\alpha_{CG}=\num{0.50}$ 
and ${\overline \epsilon}_{es} = 0$ (purple triangles). The IDPs are 
ordered from shortest to longest (left to right).} \end{figure*}

\begin{figure*}
\begin{centering}
\includegraphics[width=3in]{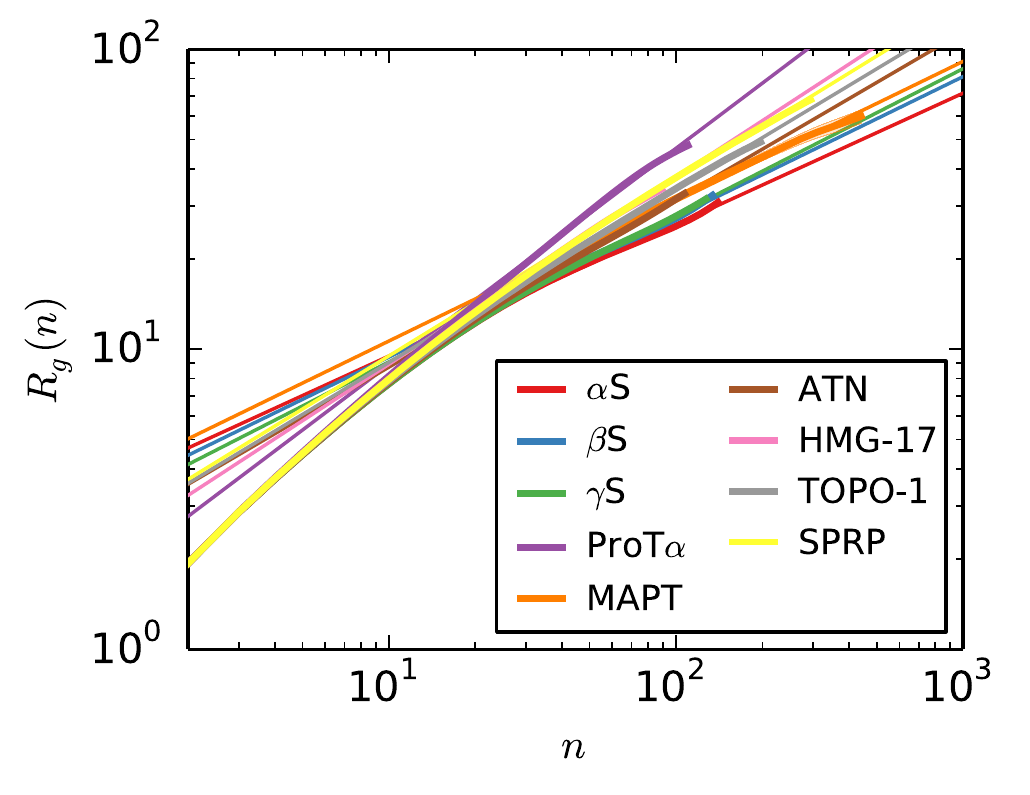}\includegraphics[width=3in]{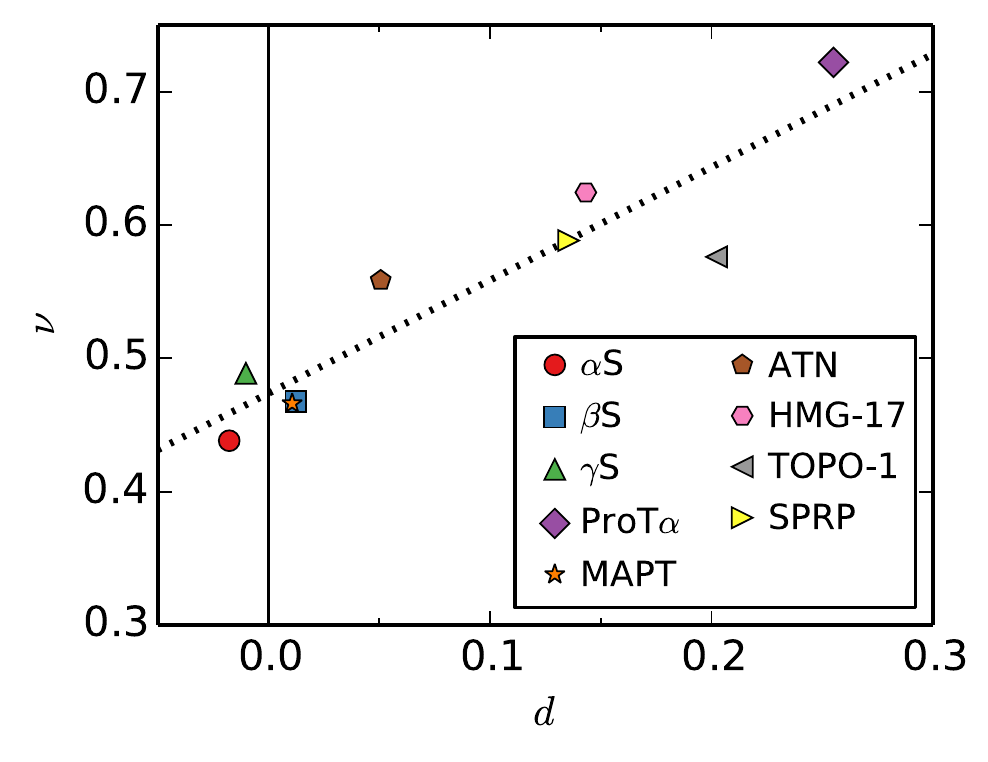}
\par\end{centering}

\protect\caption{\label{fig:Rg-scaling}(left) Radius of gyration 
$R_g(n)$ (thick lines) versus chemical distance $n$ along the chain for 
several IDPs with $N \ge 90$ so that $R_{g}(n)$ is in the power-law 
scaling regime. Power-law fits of the data to $R_g = R^0 n^{\nu}$ for 
$n > 20$ are shown as thin lines. The error in $R_{g}$ is comparable to 
the line thickness. (right) Power-law scaling exponent $\nu$ as a 
function of the distance $d$ from the dividing line between folded and 
intrinsically disordered proteins (Fig.~\ref{fig:qvh}). The dotted 
line follows $\nu=0.47+0.85 d$.} \end{figure*}

\subsection{Scaling exponents}
\label{se}

The charge-hydrophobicity plane~\citep{Uversky2000} is a common rubric 
for differentiating natively folded and intrinsically disordered 
proteins. In Fig.~\ref{fig:qvh}, we plot the absolute value of the 
electric charge per residue $Q=N_p^{-1} |\sum_{i=1}^{N_p} Q_i|$ 
versus the hydrophobicity per residue $H = N_p^{-1} 
\sum_{i=1}^{N_p} \epsilon_i$ (using the shifted and normalized Monera 
hydrophobicity scale) for many known IDPs and folded proteins.
We highlight $10$ specific proteins in Fig.~\ref{fig:qvh}: 
$\alpha$S, $\beta$S, $\gamma$S, ProT$\alpha$, high mobility 
anti-termination protein N (ATN), MAPT, non-histone chromosomal protein 
(HMG-17), DNA topoisomerase 1 (TOPO-1), basic salivary proline-rich 
protein 4 (SPRP), and lysosyme C (LYZ). The majority of IDPs occur 
above the line $Q=2.785 H -1.151$, while natively folded proteins occur 
below the line. For example, the IDPs ProT$\alpha$ and HMG-17 occur 
significantly above the line, while the folded protein lysozyme C is 
well below the line. However, the synucleins and MAPT occur close to 
the dividing line between folded and intrinsically disordered proteins. 
In fact, $\alpha$S and $\gamma$S are on the folded-protein 
side of the dividing line along with several other IDPs, and thus the 
dividing line is somewhat `fuzzy'.

We seek to identify physical quantities that are able to distinguish 
the behavior of different IDPs. In this section, we employ the CG model 
to measure the radius of gyration $R_g$ as a function of chemical 
distance $n$ along the chain

\begin{equation}
R_{g}(n)=\frac{1}{N_p-n+1} \sum_{i=1}^{N_p-n} \left\langle R_{g}(i, i+n-1)\right\rangle_{t},
\end{equation}

where $\left\langle . \right\rangle_{t}$ denotes a time average,

\begin{equation}
R_{g}(i,j)=\sqrt{\frac{1}{j-i+1}\sum_{k=i}^{j}\left( {\vec r}_k-\left\langle {\vec r_k} \right\rangle \right)^{2}},
\end{equation}

and

\begin{equation}
\left\langle {\vec r_k} \right\rangle = \frac{1}{j-i+1}\sum_{k=i}^{j} \vec r_k,
\end{equation}

for proteins over a broad range of the charge-hydrophobicity plane. In 
Fig.~\ref{fig:rgs}, we show the radius of gyration $R_{g}(N_{p})$ for 
seven IDPs (HMG-17, ATN, ProT$\alpha$, the synuclein family, and MAPT) 
ordered from shortest to longest. We find that the CG model with the 
optimal $\alpha_{CG}$ is able to recapitulate the experimental values 
of the radius of gyration for these IDPs to within approximately 
$10$\%. We also show that the predicted $R_g(N_{p})$ from the CG model 
without electrostatics, ${\overline \epsilon}_{es} = 0$, matches the 
experimental values for these IDPs, except for ProT$\alpha$ and MAPT. 
Additionally, the electrostatics-only model ($\alpha_{CG}=0$) is not 
able to recapitulate the experimental $R_{g}(N_{p})$ for most of these 
IDPs.

We show the dependence of $R_g(n)$ on the chemical distance $n$ along 
the chain for eight IDPs in Fig.~\ref{fig:Rg-scaling} (left). For large 
$n$, $R_g$ displays power-law scaling, $R_g = R_{g}^0 n^{\nu}$, where 
the scaling exponent $\nu$ varies from $\approx 0.5$ to $0.7$ as shown 
in Fig.~\ref{fig:Rg-scaling} (right). We find that there is a strong 
correlation between $\nu$ and the distance $d$ of the protein from the 
dividing line between IDPs and folded proteins in the 
charge-hydrophobicity plane. For proteins near the dividing line, the 
exponent $\nu \approx 0.5$ shows ideal scaling, while swelling of the 
chain increases linearly with distance from the dividing line.

\section{Conclusions\label{sec:Conclusions}}

We developed a coarse-grained representation of intrinsically
disordered proteins (IDPs) that includes steric, attractive
hydrophobic, and screened electrostatic interactions between spherical
residues. The CG model is calibrated to recapitulate a large set of
experimental measurements of FRET efficiencies for $5$ IDPs and $36$
pairs of residues. We then performed Langevin dynamics simulations of
the calibrated CG model to calculate the scaling of the radius of
gyration with the chemical distance along the chain for a larger set
of IDPs. We find a strong correlation between the scaling exponent
$\nu$ that characterizes the swelling of the IDP and its distance from
the line that separates IDPs and natively folded proteins in the
hydrophobicity and charge plane. IDPs possess ideal scaling $(\nu
\sim 0.5)$ near the dividing line, and the exponent increases linearly
with distance from the dividing line. These results suggest that
increasing the charge or decreasing hydrophobicity can have similar
effects on the swelling of IDPs. In future studies, we will employ
this simple, robust CG model to study the association and aggregation
dynamics of tens to hundreds of IDPs and address such questions as: 1)
Is the single-chain model for IDPs able to capture the aggregation of
multiple IDPs, 2) Does $\beta$-sheet order form spontaneously in
clusters of IDPs, and if so, 3) what is the critical nucleus for
$\beta$-sheet order?

\section*{Acknowledgments\label{sec:Acknowledgments}}

This research was supported by the National Science Foundation (NSF)
under Grant Nos. DMR-1006537 (C.O.) and PHY-1019147 (W.S.) and the 
Raymond and Beverly Sackler Institute for Biological, Physical, 
and Engineering Sciences (C.O.). This work also
benefited from the facilities and staff of the Yale University Faculty
of Arts and Sciences High Performance Computing Center and NSF Grant
No. CNS-0821132 that partially funded acquisition of the computational
facilities.

\bibliographystyle{aipauth4-1}
\bibliography{bib_draft}

\begin{thebibliography}{52}%
\makeatletter
\providecommand \@ifxundefined [1]{%
 \@ifx{#1\undefined}
}%
\providecommand \@ifnum [1]{%
 \ifnum #1\expandafter \@firstoftwo
 \else \expandafter \@secondoftwo
 \fi
}%
\providecommand \@ifx [1]{%
 \ifx #1\expandafter \@firstoftwo
 \else \expandafter \@secondoftwo
 \fi
}%
\providecommand \natexlab [1]{#1}%
\providecommand \enquote  [1]{``#1''}%
\providecommand \bibnamefont  [1]{#1}%
\providecommand \bibfnamefont [1]{#1}%
\providecommand \citenamefont [1]{#1}%
\providecommand \href@noop [0]{\@secondoftwo}%
\providecommand \href [0]{\begingroup \@sanitize@url \@href}%
\providecommand \@href[1]{\@@startlink{#1}\@@href}%
\providecommand \@@href[1]{\endgroup#1\@@endlink}%
\providecommand \@sanitize@url [0]{\catcode `\\12\catcode `\$12\catcode
  `\&12\catcode `\#12\catcode `\^12\catcode `\_12\catcode `\%12\relax}%
\providecommand \@@startlink[1]{}%
\providecommand \@@endlink[0]{}%
\providecommand \url  [0]{\begingroup\@sanitize@url \@url }%
\providecommand \@url [1]{\endgroup\@href {#1}{\urlprefix }}%
\providecommand \urlprefix  [0]{URL }%
\providecommand \Eprint [0]{\href }%
\providecommand \doibase [0]{http://dx.doi.org/}%
\providecommand \selectlanguage [0]{\@gobble}%
\providecommand \bibinfo  [0]{\@secondoftwo}%
\providecommand \bibfield  [0]{\@secondoftwo}%
\providecommand \translation [1]{[#1]}%
\providecommand \BibitemOpen [0]{}%
\providecommand \bibitemStop [0]{}%
\providecommand \bibitemNoStop [0]{.\EOS\space}%
\providecommand \EOS [0]{\spacefactor3000\relax}%
\providecommand \BibitemShut  [1]{\csname bibitem#1\endcsname}%
\let\auto@bib@innerbib\@empty
\bibitem [{\citenamefont {Abercrombie}\ \emph {et~al.}(1978)\citenamefont
  {Abercrombie}, \citenamefont {Kneale}, \citenamefont {Crane-Robinson},
  \citenamefont {Bradbury}, \citenamefont {Goodwin}, \citenamefont {Walker},\
  and\ \citenamefont {Johns}}]{Abercrombie1978}%
  \BibitemOpen
  \bibfield  {author} {\bibinfo {author} {\bibnamefont {Abercrombie},
  \bibfnamefont {B.~D.}}, \bibinfo {author} {\bibnamefont {Kneale},
  \bibfnamefont {G.~G.}}, \bibinfo {author} {\bibnamefont {Crane-Robinson},
  \bibfnamefont {C.}}, \bibinfo {author} {\bibnamefont {Bradbury},
  \bibfnamefont {E.~M.}}, \bibinfo {author} {\bibnamefont {Goodwin},
  \bibfnamefont {G.~H.}}, \bibinfo {author} {\bibnamefont {Walker},
  \bibfnamefont {J.~M.}}, \ and\ \bibinfo {author} {\bibnamefont {Johns},
  \bibfnamefont {E.~W.}},\ }\href@noop {} {\bibfield  {journal} {\bibinfo
  {journal} {European journal of biochemistry / FEBS}\ }\textbf {\bibinfo
  {volume} {84}},\ \bibinfo {pages} {173} (\bibinfo {year} {1978})}\BibitemShut
  {NoStop}%
\bibitem [{\citenamefont {Allen}\ and\ \citenamefont
  {Tildesley}(1989)}]{Allen1989}%
  \BibitemOpen
  \bibfield  {author} {\bibinfo {author} {\bibnamefont {Allen}, \bibfnamefont
  {M.~P.}}\ and\ \bibinfo {author} {\bibnamefont {Tildesley}, \bibfnamefont
  {D.~J.}},\ }\href@noop {} {\emph {\bibinfo {title} {{Computer simulation of
  liquids}}}}\ (\bibinfo  {publisher} {Oxford University Press},\ \bibinfo
  {year} {1989})\BibitemShut {NoStop}%
\bibitem [{\citenamefont {Brooks}\ \emph {et~al.}(2009)\citenamefont {Brooks},
  \citenamefont {Brooks}, \citenamefont {MacKerell}, \citenamefont {Nilsson},
  \citenamefont {Petrella}, \citenamefont {Roux}, \citenamefont {Won},
  \citenamefont {Archontis}, \citenamefont {Bartels}, \citenamefont {Boresch}
  \emph {et~al.}}]{Brooks2009}%
  \BibitemOpen
  \bibfield  {author} {\bibinfo {author} {\bibnamefont {Brooks}, \bibfnamefont
  {B.~R.}}, \bibinfo {author} {\bibnamefont {Brooks}, \bibfnamefont {C.~L.}},
  \bibinfo {author} {\bibnamefont {MacKerell}, \bibfnamefont {A.~D.}}, \bibinfo
  {author} {\bibnamefont {Nilsson}, \bibfnamefont {L.}}, \bibinfo {author}
  {\bibnamefont {Petrella}, \bibfnamefont {R.~J.}}, \bibinfo {author}
  {\bibnamefont {Roux}, \bibfnamefont {B.}}, \bibinfo {author} {\bibnamefont
  {Won}, \bibfnamefont {Y.}}, \bibinfo {author} {\bibnamefont {Archontis},
  \bibfnamefont {G.}}, \bibinfo {author} {\bibnamefont {Bartels}, \bibfnamefont
  {C.}}, \bibinfo {author} {\bibnamefont {Boresch}, \bibfnamefont {S.}},  \emph
  {et~al.},\ }\href@noop {} {\bibfield  {journal} {\bibinfo  {journal} {Journal
  of computational chemistry}\ }\textbf {\bibinfo {volume} {30}},\ \bibinfo
  {pages} {1545} (\bibinfo {year} {2009})}\BibitemShut {NoStop}%
\bibitem [{\citenamefont {Castanon}\ \emph {et~al.}(1988)\citenamefont
  {Castanon}, \citenamefont {Spevak}, \citenamefont {Adolf}, \citenamefont
  {Chlebowicz-Sledziewska},\ and\ \citenamefont {Sledziewski}}]{Castanon1988}%
  \BibitemOpen
  \bibfield  {author} {\bibinfo {author} {\bibnamefont {Castanon},
  \bibfnamefont {M.}}, \bibinfo {author} {\bibnamefont {Spevak}, \bibfnamefont
  {W.}}, \bibinfo {author} {\bibnamefont {Adolf}, \bibfnamefont {G.}}, \bibinfo
  {author} {\bibnamefont {Chlebowicz-Sledziewska}, \bibfnamefont {E.}}, \ and\
  \bibinfo {author} {\bibnamefont {Sledziewski}, \bibfnamefont {A.}},\
  }\href@noop {} {\bibfield  {journal} {\bibinfo  {journal} {Gene}\ }\textbf
  {\bibinfo {volume} {66}},\ \bibinfo {pages} {223} (\bibinfo {year}
  {1988})}\BibitemShut {NoStop}%
\bibitem [{\citenamefont {Clementi}(2008)}]{Clementi2008}%
  \BibitemOpen
  \bibfield  {author} {\bibinfo {author} {\bibnamefont {Clementi},
  \bibfnamefont {C.}},\ }\href {\doibase 10.1016/j.sbi.2007.10.005} {\bibfield
  {journal} {\bibinfo  {journal} {Current opinion in structural biology}\
  }\textbf {\bibinfo {volume} {18}},\ \bibinfo {pages} {10} (\bibinfo {year}
  {2008})}\BibitemShut {NoStop}%
\bibitem [{\citenamefont {Cornette}\ \emph {et~al.}(1987)\citenamefont
  {Cornette}, \citenamefont {Cease}, \citenamefont {Margalit}, \citenamefont
  {Spouge}, \citenamefont {Berzofsky},\ and\ \citenamefont
  {DeLisi}}]{Cornette1987}%
  \BibitemOpen
  \bibfield  {author} {\bibinfo {author} {\bibnamefont {Cornette},
  \bibfnamefont {J.~L.}}, \bibinfo {author} {\bibnamefont {Cease},
  \bibfnamefont {K.~B.}}, \bibinfo {author} {\bibnamefont {Margalit},
  \bibfnamefont {H.}}, \bibinfo {author} {\bibnamefont {Spouge}, \bibfnamefont
  {J.~L.}}, \bibinfo {author} {\bibnamefont {Berzofsky}, \bibfnamefont
  {J.~A.}}, \ and\ \bibinfo {author} {\bibnamefont {DeLisi}, \bibfnamefont
  {C.}},\ }\href
  {http://www.sciencedirect.com/science/article/pii/0022283687901896}
  {\bibfield  {journal} {\bibinfo  {journal} {Journal of molecular biology}\
  }\textbf {\bibinfo {volume} {195}},\ \bibinfo {pages} {659} (\bibinfo {year}
  {1987})}\BibitemShut {NoStop}%
\bibitem [{\citenamefont {Dedmon}\ \emph {et~al.}(2005)\citenamefont {Dedmon},
  \citenamefont {Lindorff-Larsen}, \citenamefont {Christodoulou}, \citenamefont
  {Vendruscolo},\ and\ \citenamefont {Dobson}}]{Dedmon2005}%
  \BibitemOpen
  \bibfield  {author} {\bibinfo {author} {\bibnamefont {Dedmon}, \bibfnamefont
  {M.~M.}}, \bibinfo {author} {\bibnamefont {Lindorff-Larsen}, \bibfnamefont
  {K.}}, \bibinfo {author} {\bibnamefont {Christodoulou}, \bibfnamefont {J.}},
  \bibinfo {author} {\bibnamefont {Vendruscolo}, \bibfnamefont {M.}}, \ and\
  \bibinfo {author} {\bibnamefont {Dobson}, \bibfnamefont {C.~M.}},\ }\href
  {\doibase 10.1021/ja044834j} {\bibfield  {journal} {\bibinfo  {journal}
  {Journal of the American Chemical Society}\ }\textbf {\bibinfo {volume}
  {127}},\ \bibinfo {pages} {476} (\bibinfo {year} {2005})}\BibitemShut
  {NoStop}%
\bibitem [{\citenamefont {Dobson}(2003)}]{Dobson2003}%
  \BibitemOpen
  \bibfield  {author} {\bibinfo {author} {\bibnamefont {Dobson}, \bibfnamefont
  {C.~M.}},\ }\href
  {http://www.nature.com/nature/journal/v426/n6968/abs/nature02261.html}
  {\bibfield  {journal} {\bibinfo  {journal} {Nature}\ }\textbf {\bibinfo
  {volume} {426}},\ \bibinfo {pages} {884} (\bibinfo {year}
  {2003})}\BibitemShut {NoStop}%
\bibitem [{\citenamefont {Ducas}\ and\ \citenamefont
  {Rhoades}(2014)}]{ducas2014}%
  \BibitemOpen
  \bibfield  {author} {\bibinfo {author} {\bibnamefont {Ducas}, \bibfnamefont
  {V.~C.}}\ and\ \bibinfo {author} {\bibnamefont {Rhoades}, \bibfnamefont
  {E.}},\ }\href@noop {} {\bibfield  {journal} {\bibinfo  {journal} {PLOS ONE}\
  }\textbf {\bibinfo {volume} {9}},\ \bibinfo {pages} {e86983} (\bibinfo {year}
  {2014})}\BibitemShut {NoStop}%
\bibitem [{\citenamefont {Dunker}\ \emph {et~al.}(2001)\citenamefont {Dunker},
  \citenamefont {Lawson}, \citenamefont {Brown}, \citenamefont {Williams},
  \citenamefont {Romero}, \citenamefont {Oh}, \citenamefont {Oldfield},
  \citenamefont {Campen}, \citenamefont {Ratliff}, \citenamefont {Hipps},
  \citenamefont {Ausio}, \citenamefont {Nissen}, \citenamefont {Reeves},
  \citenamefont {Kang}, \citenamefont {Kissinger}, \citenamefont {Bailey},
  \citenamefont {Griswold}, \citenamefont {Chiu}, \citenamefont {Garner},\ and\
  \citenamefont {Obradovic}}]{Dunker2001}%
  \BibitemOpen
  \bibfield  {author} {\bibinfo {author} {\bibnamefont {Dunker}, \bibfnamefont
  {A.}}, \bibinfo {author} {\bibnamefont {Lawson}, \bibfnamefont {J.}},
  \bibinfo {author} {\bibnamefont {Brown}, \bibfnamefont {C.~J.}}, \bibinfo
  {author} {\bibnamefont {Williams}, \bibfnamefont {R.~M.}}, \bibinfo {author}
  {\bibnamefont {Romero}, \bibfnamefont {P.}}, \bibinfo {author} {\bibnamefont
  {Oh}, \bibfnamefont {J.~S.}}, \bibinfo {author} {\bibnamefont {Oldfield},
  \bibfnamefont {C.~J.}}, \bibinfo {author} {\bibnamefont {Campen},
  \bibfnamefont {A.~M.}}, \bibinfo {author} {\bibnamefont {Ratliff},
  \bibfnamefont {C.~M.}}, \bibinfo {author} {\bibnamefont {Hipps},
  \bibfnamefont {K.~W.}}, \bibinfo {author} {\bibnamefont {Ausio},
  \bibfnamefont {J.}}, \bibinfo {author} {\bibnamefont {Nissen}, \bibfnamefont
  {M.~S.}}, \bibinfo {author} {\bibnamefont {Reeves}, \bibfnamefont {R.}},
  \bibinfo {author} {\bibnamefont {Kang}, \bibfnamefont {C.}}, \bibinfo
  {author} {\bibnamefont {Kissinger}, \bibfnamefont {C.~R.}}, \bibinfo {author}
  {\bibnamefont {Bailey}, \bibfnamefont {R.~W.}}, \bibinfo {author}
  {\bibnamefont {Griswold}, \bibfnamefont {M.~D.}}, \bibinfo {author}
  {\bibnamefont {Chiu}, \bibfnamefont {W.}}, \bibinfo {author} {\bibnamefont
  {Garner}, \bibfnamefont {E.~C.}}, \ and\ \bibinfo {author} {\bibnamefont
  {Obradovic}, \bibfnamefont {Z.}},\ }\href@noop {} {\bibfield  {journal}
  {\bibinfo  {journal} {Journal of Molecular Graphics and Modelling}\ }\textbf
  {\bibinfo {volume} {19}},\ \bibinfo {pages} {26} (\bibinfo {year}
  {2001})}\BibitemShut {NoStop}%
\bibitem [{\citenamefont {Dyson}\ and\ \citenamefont
  {Wright}(2005)}]{Dyson2005}%
  \BibitemOpen
  \bibfield  {author} {\bibinfo {author} {\bibnamefont {Dyson}, \bibfnamefont
  {H.~J.}}\ and\ \bibinfo {author} {\bibnamefont {Wright}, \bibfnamefont
  {P.~E.}},\ }\href {\doibase 10.1038/nrm1589} {\bibfield  {journal} {\bibinfo
  {journal} {Nature reviews. Molecular cell biology}\ }\textbf {\bibinfo
  {volume} {6}},\ \bibinfo {pages} {197} (\bibinfo {year} {2005})}\BibitemShut
  {NoStop}%
\bibitem [{\citenamefont {Eisenberg}\ \emph {et~al.}(1984)\citenamefont
  {Eisenberg}, \citenamefont {Schwarz}, \citenamefont {Komaromy},\ and\
  \citenamefont {Wall}}]{Eisenberg1984}%
  \BibitemOpen
  \bibfield  {author} {\bibinfo {author} {\bibnamefont {Eisenberg},
  \bibfnamefont {D.}}, \bibinfo {author} {\bibnamefont {Schwarz}, \bibfnamefont
  {E.}}, \bibinfo {author} {\bibnamefont {Komaromy}, \bibfnamefont {M.}}, \
  and\ \bibinfo {author} {\bibnamefont {Wall}, \bibfnamefont {R.}},\
  }\href@noop {} {\bibfield  {journal} {\bibinfo  {journal} {Journal of
  molecular biology}\ }\textbf {\bibinfo {volume} {179}},\ \bibinfo {pages}
  {125} (\bibinfo {year} {1984})}\BibitemShut {NoStop}%
\bibitem [{\citenamefont {Eschenfeldt}\ and\ \citenamefont
  {Berger}(1986)}]{Eschenfeldt1986}%
  \BibitemOpen
  \bibfield  {author} {\bibinfo {author} {\bibnamefont {Eschenfeldt},
  \bibfnamefont {W.~H.}}\ and\ \bibinfo {author} {\bibnamefont {Berger},
  \bibfnamefont {S.~L.}},\ }\href
  {http://www.pnas.org/content/83/24/9403.short} {\bibfield  {journal}
  {\bibinfo  {journal} {Proceedings of the National Academy of Sciences}\
  }\textbf {\bibinfo {volume} {83}},\ \bibinfo {pages} {9403} (\bibinfo {year}
  {1986})}\BibitemShut {NoStop}%
\bibitem [{\citenamefont {Gast}\ \emph {et~al.}(1995)\citenamefont {Gast},
  \citenamefont {Damaschun}, \citenamefont {Eckert}, \citenamefont
  {Schulze-Forster}, \citenamefont {Maurer}, \citenamefont {Mueller-Frohne},
  \citenamefont {Zirwer}, \citenamefont {Czarnecki},\ and\ \citenamefont
  {Damaschun}}]{Gast1995}%
  \BibitemOpen
  \bibfield  {author} {\bibinfo {author} {\bibnamefont {Gast}, \bibfnamefont
  {K.}}, \bibinfo {author} {\bibnamefont {Damaschun}, \bibfnamefont {H.}},
  \bibinfo {author} {\bibnamefont {Eckert}, \bibfnamefont {K.}}, \bibinfo
  {author} {\bibnamefont {Schulze-Forster}, \bibfnamefont {K.}}, \bibinfo
  {author} {\bibnamefont {Maurer}, \bibfnamefont {H.~R.}}, \bibinfo {author}
  {\bibnamefont {Mueller-Frohne}, \bibfnamefont {M.}}, \bibinfo {author}
  {\bibnamefont {Zirwer}, \bibfnamefont {D.}}, \bibinfo {author} {\bibnamefont
  {Czarnecki}, \bibfnamefont {J.}}, \ and\ \bibinfo {author} {\bibnamefont
  {Damaschun}, \bibfnamefont {G.}},\ }\href@noop {} {\bibfield  {journal}
  {\bibinfo  {journal} {Biochemistry}\ }\textbf {\bibinfo {volume} {34}},\
  \bibinfo {pages} {13211} (\bibinfo {year} {1995})}\BibitemShut {NoStop}%
\bibitem [{\citenamefont {Giehm}\ \emph {et~al.}(2011)\citenamefont {Giehm},
  \citenamefont {Svergun}, \citenamefont {Otzen},\ and\ \citenamefont
  {Vestergaard}}]{giehm2011}%
  \BibitemOpen
  \bibfield  {author} {\bibinfo {author} {\bibnamefont {Giehm}, \bibfnamefont
  {L.}}, \bibinfo {author} {\bibnamefont {Svergun}, \bibfnamefont {D.~I.}},
  \bibinfo {author} {\bibnamefont {Otzen}, \bibfnamefont {D.~E.}}, \ and\
  \bibinfo {author} {\bibnamefont {Vestergaard}, \bibfnamefont {B.}},\ }\href
  {\doibase 10.1073/pnas.1013225108} {\bibfield  {journal} {\bibinfo  {journal}
  {Proceedings of the National Academy of Sciences}\ }\textbf {\bibinfo
  {volume} {108}},\ \bibinfo {pages} {3246} (\bibinfo {year}
  {2011})}\BibitemShut {NoStop}%
\bibitem [{\citenamefont {Goedert}\ \emph {et~al.}(1989)\citenamefont
  {Goedert}, \citenamefont {Spillantini}, \citenamefont {Jakes}, \citenamefont
  {Rutherford},\ and\ \citenamefont {Crowther}}]{Goedert1989}%
  \BibitemOpen
  \bibfield  {author} {\bibinfo {author} {\bibnamefont {Goedert}, \bibfnamefont
  {M.}}, \bibinfo {author} {\bibnamefont {Spillantini}, \bibfnamefont {M.}},
  \bibinfo {author} {\bibnamefont {Jakes}, \bibfnamefont {R.}}, \bibinfo
  {author} {\bibnamefont {Rutherford}, \bibfnamefont {D.}}, \ and\ \bibinfo
  {author} {\bibnamefont {Crowther}, \bibfnamefont {R.}},\ }\href@noop {}
  {\bibfield  {journal} {\bibinfo  {journal} {Neuron}\ }\textbf {\bibinfo
  {volume} {3}},\ \bibinfo {pages} {519} (\bibinfo {year} {1989})}\BibitemShut
  {NoStop}%
\bibitem [{\citenamefont {Goedert}\ \emph {et~al.}(1988)\citenamefont
  {Goedert}, \citenamefont {Wischik}, \citenamefont {Crowther}, \citenamefont
  {Walker},\ and\ \citenamefont {Klug}}]{Goedert1988}%
  \BibitemOpen
  \bibfield  {author} {\bibinfo {author} {\bibnamefont {Goedert}, \bibfnamefont
  {M.}}, \bibinfo {author} {\bibnamefont {Wischik}, \bibfnamefont {C.}},
  \bibinfo {author} {\bibnamefont {Crowther}, \bibfnamefont {R.}}, \bibinfo
  {author} {\bibnamefont {Walker}, \bibfnamefont {J.}}, \ and\ \bibinfo
  {author} {\bibnamefont {Klug}, \bibfnamefont {A.}},\ }\href@noop {}
  {\bibfield  {journal} {\bibinfo  {journal} {Proceedings of the National
  Academy of Sciences}\ }\textbf {\bibinfo {volume} {85}},\ \bibinfo {pages}
  {4051} (\bibinfo {year} {1988})}\BibitemShut {NoStop}%
\bibitem [{\citenamefont {Hashimoto}\ \emph {et~al.}(2001)\citenamefont
  {Hashimoto}, \citenamefont {Rockenstein}, \citenamefont {Mante},
  \citenamefont {Mallory},\ and\ \citenamefont {Masliah}}]{Hashimoto2001}%
  \BibitemOpen
  \bibfield  {author} {\bibinfo {author} {\bibnamefont {Hashimoto},
  \bibfnamefont {M.}}, \bibinfo {author} {\bibnamefont {Rockenstein},
  \bibfnamefont {E.}}, \bibinfo {author} {\bibnamefont {Mante}, \bibfnamefont
  {M.}}, \bibinfo {author} {\bibnamefont {Mallory}, \bibfnamefont {M.}}, \ and\
  \bibinfo {author} {\bibnamefont {Masliah}, \bibfnamefont {E.}},\ }\href@noop
  {} {\bibfield  {journal} {\bibinfo  {journal} {Neuron}\ }\textbf {\bibinfo
  {volume} {32}},\ \bibinfo {pages} {213} (\bibinfo {year} {2001})}\BibitemShut
  {NoStop}%
\bibitem [{\citenamefont {Hofmann}\ \emph {et~al.}(2012)\citenamefont
  {Hofmann}, \citenamefont {Soranno}, \citenamefont {Borgia}, \citenamefont
  {Gast}, \citenamefont {Nettels},\ and\ \citenamefont
  {Schuler}}]{Hofmann2012}%
  \BibitemOpen
  \bibfield  {author} {\bibinfo {author} {\bibnamefont {Hofmann}, \bibfnamefont
  {H.}}, \bibinfo {author} {\bibnamefont {Soranno}, \bibfnamefont {A.}},
  \bibinfo {author} {\bibnamefont {Borgia}, \bibfnamefont {A.}}, \bibinfo
  {author} {\bibnamefont {Gast}, \bibfnamefont {K.}}, \bibinfo {author}
  {\bibnamefont {Nettels}, \bibfnamefont {D.}}, \ and\ \bibinfo {author}
  {\bibnamefont {Schuler}, \bibfnamefont {B.}},\ }\href@noop {} {\bibfield
  {journal} {\bibinfo  {journal} {Proceedings of the National Academy of
  Sciences of the United States of America}\ }\textbf {\bibinfo {volume}
  {109}},\ \bibinfo {pages} {16155} (\bibinfo {year} {2012})}\BibitemShut
  {NoStop}%
\bibitem [{\citenamefont {Jakes}, \citenamefont {Spillantini},\ and\
  \citenamefont {Goedert}(1994)}]{Jakes1994}%
  \BibitemOpen
  \bibfield  {author} {\bibinfo {author} {\bibnamefont {Jakes}, \bibfnamefont
  {R.}}, \bibinfo {author} {\bibnamefont {Spillantini}, \bibfnamefont {M.~G.}},
  \ and\ \bibinfo {author} {\bibnamefont {Goedert}, \bibfnamefont {M.}},\
  }\href@noop {} {\bibfield  {journal} {\bibinfo  {journal} {FEBS letters}\
  }\textbf {\bibinfo {volume} {345}},\ \bibinfo {pages} {27} (\bibinfo {year}
  {1994})}\BibitemShut {NoStop}%
\bibitem [{\citenamefont {Ji}\ \emph {et~al.}(1997)\citenamefont {Ji},
  \citenamefont {Liu}, \citenamefont {Jia}, \citenamefont {Liu}, \citenamefont
  {Xiao}, \citenamefont {Joseph}, \citenamefont {Rosen},\ and\ \citenamefont
  {Sm}}]{Ji1997}%
  \BibitemOpen
  \bibfield  {author} {\bibinfo {author} {\bibnamefont {Ji}, \bibfnamefont
  {H.}}, \bibinfo {author} {\bibnamefont {Liu}, \bibfnamefont {Y.~E.}},
  \bibinfo {author} {\bibnamefont {Jia}, \bibfnamefont {T.}}, \bibinfo {author}
  {\bibnamefont {Liu}, \bibfnamefont {Y.~E.}}, \bibinfo {author} {\bibnamefont
  {Xiao}, \bibfnamefont {G.}}, \bibinfo {author} {\bibnamefont {Joseph},
  \bibfnamefont {K.}}, \bibinfo {author} {\bibnamefont {Rosen}, \bibfnamefont
  {C.}}, \ and\ \bibinfo {author} {\bibnamefont {Sm}, \bibfnamefont {Y.~E.}},\
  }\href@noop {} {\bibfield  {journal} {\bibinfo  {journal} {Cancer Res}\
  }\textbf {\bibinfo {volume} {57}},\ \bibinfo {pages} {759} (\bibinfo {year}
  {1997})}\BibitemShut {NoStop}%
\bibitem [{\citenamefont {Johansen}\ \emph {et~al.}(2011)\citenamefont
  {Johansen}, \citenamefont {Jeffries}, \citenamefont {Hammouda}, \citenamefont
  {Trewhella},\ and\ \citenamefont {Goldenberg}}]{Johansen2011}%
  \BibitemOpen
  \bibfield  {author} {\bibinfo {author} {\bibnamefont {Johansen},
  \bibfnamefont {D.}}, \bibinfo {author} {\bibnamefont {Jeffries},
  \bibfnamefont {C.~M.~J.}}, \bibinfo {author} {\bibnamefont {Hammouda},
  \bibfnamefont {B.}}, \bibinfo {author} {\bibnamefont {Trewhella},
  \bibfnamefont {J.}}, \ and\ \bibinfo {author} {\bibnamefont {Goldenberg},
  \bibfnamefont {D.~P.}},\ }\href {\doibase 10.1016/j.bpj.2011.01.020}
  {\bibfield  {journal} {\bibinfo  {journal} {Biophysical journal}\ }\textbf
  {\bibinfo {volume} {100}},\ \bibinfo {pages} {1120} (\bibinfo {year}
  {2011})}\BibitemShut {NoStop}%
\bibitem [{\citenamefont {Kyte}\ and\ \citenamefont
  {Doolittle}(1982)}]{Kyte1982}%
  \BibitemOpen
  \bibfield  {author} {\bibinfo {author} {\bibnamefont {Kyte}, \bibfnamefont
  {J.}}\ and\ \bibinfo {author} {\bibnamefont {Doolittle}, \bibfnamefont
  {R.~F.}},\ }\href@noop {} {\bibfield  {journal} {\bibinfo  {journal} {Journal
  of molecular biology}\ }\textbf {\bibinfo {volume} {157}},\ \bibinfo {pages}
  {105} (\bibinfo {year} {1982})}\BibitemShut {NoStop}%
\bibitem [{\citenamefont {Li}, \citenamefont {Uversky},\ and\ \citenamefont
  {Fink}(2002)}]{li2002}%
  \BibitemOpen
  \bibfield  {author} {\bibinfo {author} {\bibnamefont {Li}, \bibfnamefont
  {J.}}, \bibinfo {author} {\bibnamefont {Uversky}, \bibfnamefont {V.~N.}}, \
  and\ \bibinfo {author} {\bibnamefont {Fink}, \bibfnamefont {A.~L.}},\ }\href
  {\doibase 10.1016/S0161-813X(02)00066-9} {\bibfield  {journal} {\bibinfo
  {journal} {{NeuroToxicology}}\ }\textbf {\bibinfo {volume} {23}},\ \bibinfo
  {pages} {553} (\bibinfo {year} {2002})}\BibitemShut {NoStop}%
\bibitem [{\citenamefont {Mao}\ \emph {et~al.}(2010)\citenamefont {Mao},
  \citenamefont {Crick}, \citenamefont {Vitalis}, \citenamefont {Chicoine},\
  and\ \citenamefont {Pappu}}]{Mao2010}%
  \BibitemOpen
  \bibfield  {author} {\bibinfo {author} {\bibnamefont {Mao}, \bibfnamefont
  {A.~H.}}, \bibinfo {author} {\bibnamefont {Crick}, \bibfnamefont {S.~L.}},
  \bibinfo {author} {\bibnamefont {Vitalis}, \bibfnamefont {A.}}, \bibinfo
  {author} {\bibnamefont {Chicoine}, \bibfnamefont {C.~L.}}, \ and\ \bibinfo
  {author} {\bibnamefont {Pappu}, \bibfnamefont {R.~V.}},\ }\href {\doibase
  10.1073/pnas.0911107107} {\bibfield  {journal} {\bibinfo  {journal}
  {Proceedings of the National Academy of Sciences of the United States of
  America}\ }\textbf {\bibinfo {volume} {107}},\ \bibinfo {pages} {8183}
  (\bibinfo {year} {2010})}\BibitemShut {NoStop}%
\bibitem [{\citenamefont {Mittag}\ and\ \citenamefont
  {Forman-Kay}(2007)}]{Mittag2007}%
  \BibitemOpen
  \bibfield  {author} {\bibinfo {author} {\bibnamefont {Mittag}, \bibfnamefont
  {T.}}\ and\ \bibinfo {author} {\bibnamefont {Forman-Kay}, \bibfnamefont
  {J.~D.}},\ }\href {\doibase 10.1016/j.sbi.2007.01.009} {\bibfield  {journal}
  {\bibinfo  {journal} {Current opinion in structural biology}\ }\textbf
  {\bibinfo {volume} {17}},\ \bibinfo {pages} {3} (\bibinfo {year}
  {2007})}\BibitemShut {NoStop}%
\bibitem [{\citenamefont {Miyazawa}\ and\ \citenamefont
  {Jernigan}(1985)}]{Miyazawa1985}%
  \BibitemOpen
  \bibfield  {author} {\bibinfo {author} {\bibnamefont {Miyazawa},
  \bibfnamefont {S.}}\ and\ \bibinfo {author} {\bibnamefont {Jernigan},
  \bibfnamefont {R.~L.}},\ }\href
  {http://pubs.acs.org/doi/abs/10.1021/ma00145a039} {\bibfield  {journal}
  {\bibinfo  {journal} {Macromolecules}\ }\textbf {\bibinfo {volume} {18}},\
  \bibinfo {pages} {534} (\bibinfo {year} {1985})}\BibitemShut {NoStop}%
\bibitem [{\citenamefont {Monera}\ \emph {et~al.}(1995)\citenamefont {Monera},
  \citenamefont {Sereda}, \citenamefont {Zhou}, \citenamefont {Kay},\ and\
  \citenamefont {Hodges}}]{Monera1995}%
  \BibitemOpen
  \bibfield  {author} {\bibinfo {author} {\bibnamefont {Monera}, \bibfnamefont
  {O.~D.}}, \bibinfo {author} {\bibnamefont {Sereda}, \bibfnamefont {T.~J.}},
  \bibinfo {author} {\bibnamefont {Zhou}, \bibfnamefont {N.~E.}}, \bibinfo
  {author} {\bibnamefont {Kay}, \bibfnamefont {C.~M.}}, \ and\ \bibinfo
  {author} {\bibnamefont {Hodges}, \bibfnamefont {R.~S.}},\ }\href {\doibase
  10.1002/psc.310010507} {\bibfield  {journal} {\bibinfo  {journal} {Journal of
  peptide science : an official publication of the European Peptide Society}\
  }\textbf {\bibinfo {volume} {1}},\ \bibinfo {pages} {319} (\bibinfo {year}
  {1995})}\BibitemShut {NoStop}%
\bibitem [{\citenamefont {Morar}\ \emph {et~al.}(2001)\citenamefont {Morar},
  \citenamefont {Olteanu}, \citenamefont {Young},\ and\ \citenamefont
  {Pielak}}]{Morar2001}%
  \BibitemOpen
  \bibfield  {author} {\bibinfo {author} {\bibnamefont {Morar}, \bibfnamefont
  {A.~S.}}, \bibinfo {author} {\bibnamefont {Olteanu}, \bibfnamefont {A.}},
  \bibinfo {author} {\bibnamefont {Young}, \bibfnamefont {G.~B.}}, \ and\
  \bibinfo {author} {\bibnamefont {Pielak}, \bibfnamefont {G.~J.}},\ }\href
  {\doibase 10.1110/ps.24301} {\bibfield  {journal} {\bibinfo  {journal}
  {Protein Science}\ }\textbf {\bibinfo {volume} {10}},\ \bibinfo {pages}
  {2195} (\bibinfo {year} {2001})}\BibitemShut {NoStop}%
\bibitem [{\citenamefont {Müller-Späth}\ \emph {et~al.}(2010)\citenamefont
  {Müller-Späth}, \citenamefont {Soranno}, \citenamefont {Hirschfeld},
  \citenamefont {Hofmann}, \citenamefont {Rüegger}, \citenamefont {Reymond},
  \citenamefont {Nettels},\ and\ \citenamefont {Schuler}}]{MullerSpath2010}%
  \BibitemOpen
  \bibfield  {author} {\bibinfo {author} {\bibnamefont {Müller-Späth},
  \bibfnamefont {S.}}, \bibinfo {author} {\bibnamefont {Soranno}, \bibfnamefont
  {A.}}, \bibinfo {author} {\bibnamefont {Hirschfeld}, \bibfnamefont {V.}},
  \bibinfo {author} {\bibnamefont {Hofmann}, \bibfnamefont {H.}}, \bibinfo
  {author} {\bibnamefont {Rüegger}, \bibfnamefont {S.}}, \bibinfo {author}
  {\bibnamefont {Reymond}, \bibfnamefont {L.}}, \bibinfo {author} {\bibnamefont
  {Nettels}, \bibfnamefont {D.}}, \ and\ \bibinfo {author} {\bibnamefont
  {Schuler}, \bibfnamefont {B.}},\ }\href {\doibase 10.1073/pnas.1001743107}
  {\bibfield  {journal} {\bibinfo  {journal} {Proceedings of the National
  Academy of Sciences}\ }\textbf {\bibinfo {volume} {107}},\ \bibinfo {pages}
  {14609} (\bibinfo {year} {2010})}\BibitemShut {NoStop}%
\bibitem [{\citenamefont {Nath}\ \emph {et~al.}(2012)\citenamefont {Nath},
  \citenamefont {Sammalkorpi}, \citenamefont {DeWitt}, \citenamefont {Trexler},
  \citenamefont {Elbaum-Garfinkle}, \citenamefont {O'Hern},\ and\ \citenamefont
  {Rhoades}}]{nath}%
  \BibitemOpen
  \bibfield  {author} {\bibinfo {author} {\bibnamefont {Nath}, \bibfnamefont
  {A.}}, \bibinfo {author} {\bibnamefont {Sammalkorpi}, \bibfnamefont {M.}},
  \bibinfo {author} {\bibnamefont {DeWitt}, \bibfnamefont {D.}}, \bibinfo
  {author} {\bibnamefont {Trexler}, \bibfnamefont {A.}}, \bibinfo {author}
  {\bibnamefont {Elbaum-Garfinkle}, \bibfnamefont {S.}}, \bibinfo {author}
  {\bibnamefont {O'Hern}, \bibfnamefont {C.}}, \ and\ \bibinfo {author}
  {\bibnamefont {Rhoades}, \bibfnamefont {E.}},\ }\href@noop {} {\bibfield
  {journal} {\bibinfo  {journal} {Biophys. J.}\ }\textbf {\bibinfo {volume}
  {103}},\ \bibinfo {pages} {1940} (\bibinfo {year} {2012})}\BibitemShut
  {NoStop}%
\bibitem [{\citenamefont {Nath}\ \emph {et~al.}(2010)\citenamefont {Nath},
  \citenamefont {Meuvis}, \citenamefont {Hendrix}, \citenamefont {Carl},\ and\
  \citenamefont {Engelborghs}}]{Nath2010}%
  \BibitemOpen
  \bibfield  {author} {\bibinfo {author} {\bibnamefont {Nath}, \bibfnamefont
  {S.}}, \bibinfo {author} {\bibnamefont {Meuvis}, \bibfnamefont {J.}},
  \bibinfo {author} {\bibnamefont {Hendrix}, \bibfnamefont {J.}}, \bibinfo
  {author} {\bibnamefont {Carl}, \bibfnamefont {S.~A.}}, \ and\ \bibinfo
  {author} {\bibnamefont {Engelborghs}, \bibfnamefont {Y.}},\ }\href {\doibase
  http://dx.doi.org/10.1016/j.bpj.2009.12.4290} {\bibfield  {journal} {\bibinfo
   {journal} {Biophysical Journal}\ }\textbf {\bibinfo {volume} {98}},\
  \bibinfo {pages} {1302 } (\bibinfo {year} {2010})}\BibitemShut {NoStop}%
\bibitem [{\citenamefont {Oostenbrink}\ \emph {et~al.}(2004)\citenamefont
  {Oostenbrink}, \citenamefont {Villa}, \citenamefont {Mark},\ and\
  \citenamefont {van Gunsteren}}]{Oostenbrink2004}%
  \BibitemOpen
  \bibfield  {author} {\bibinfo {author} {\bibnamefont {Oostenbrink},
  \bibfnamefont {C.}}, \bibinfo {author} {\bibnamefont {Villa}, \bibfnamefont
  {A.}}, \bibinfo {author} {\bibnamefont {Mark}, \bibfnamefont {A.~E.}}, \ and\
  \bibinfo {author} {\bibnamefont {van Gunsteren}, \bibfnamefont {W.~F.}},\
  }\href {\doibase 10.1002/jcc.20090} {\bibfield  {journal} {\bibinfo
  {journal} {Journal of computational chemistry}\ }\textbf {\bibinfo {volume}
  {25}},\ \bibinfo {pages} {1656} (\bibinfo {year} {2004})}\BibitemShut
  {NoStop}%
\bibitem [{\citenamefont {Rekas}\ \emph {et~al.}(2010)\citenamefont {Rekas},
  \citenamefont {Knott}, \citenamefont {Sokolova}, \citenamefont {Barnham},
  \citenamefont {Perez}, \citenamefont {Masters}, \citenamefont {Drew},
  \citenamefont {Cappai}, \citenamefont {Curtain},\ and\ \citenamefont
  {Pham}}]{rekas2010}%
  \BibitemOpen
  \bibfield  {author} {\bibinfo {author} {\bibnamefont {Rekas}, \bibfnamefont
  {A.}}, \bibinfo {author} {\bibnamefont {Knott}, \bibfnamefont {R.}}, \bibinfo
  {author} {\bibnamefont {Sokolova}, \bibfnamefont {A.}}, \bibinfo {author}
  {\bibnamefont {Barnham}, \bibfnamefont {K.}}, \bibinfo {author} {\bibnamefont
  {Perez}, \bibfnamefont {K.}}, \bibinfo {author} {\bibnamefont {Masters},
  \bibfnamefont {C.}}, \bibinfo {author} {\bibnamefont {Drew}, \bibfnamefont
  {S.}}, \bibinfo {author} {\bibnamefont {Cappai}, \bibfnamefont {R.}},
  \bibinfo {author} {\bibnamefont {Curtain}, \bibfnamefont {C.}}, \ and\
  \bibinfo {author} {\bibnamefont {Pham}, \bibfnamefont {C.}},\ }\href
  {\doibase 10.1007/s00249-010-0595-x} {\bibfield  {journal} {\bibinfo
  {journal} {European Biophysics Journal}\ }\textbf {\bibinfo {volume} {39}},\
  \bibinfo {pages} {1407} (\bibinfo {year} {2010})}\BibitemShut {NoStop}%
\bibitem [{\citenamefont {Salmon}\ \emph {et~al.}(2010)\citenamefont {Salmon},
  \citenamefont {Nodet}, \citenamefont {Ozenne}, \citenamefont {Yin},
  \citenamefont {Jensen}, \citenamefont {Zweckstetter},\ and\ \citenamefont
  {Blackledge}}]{salmon2010}%
  \BibitemOpen
  \bibfield  {author} {\bibinfo {author} {\bibnamefont {Salmon}, \bibfnamefont
  {L.}}, \bibinfo {author} {\bibnamefont {Nodet}, \bibfnamefont {G.}}, \bibinfo
  {author} {\bibnamefont {Ozenne}, \bibfnamefont {V.}}, \bibinfo {author}
  {\bibnamefont {Yin}, \bibfnamefont {G.}}, \bibinfo {author} {\bibnamefont
  {Jensen}, \bibfnamefont {M.~R.}}, \bibinfo {author} {\bibnamefont
  {Zweckstetter}, \bibfnamefont {M.}}, \ and\ \bibinfo {author} {\bibnamefont
  {Blackledge}, \bibfnamefont {M.}},\ }\href
  {http://pubs.acs.org/doi/abs/10.1021/ja101645g} {\bibfield  {journal}
  {\bibinfo  {journal} {Journal of the American Chemical Society}\ }\textbf
  {\bibinfo {volume} {132}},\ \bibinfo {pages} {8407} (\bibinfo {year}
  {2010})}\BibitemShut {NoStop}%
\bibitem [{\citenamefont {Salomon-Ferrer}, \citenamefont {Case},\ and\
  \citenamefont {Walker}(2013)}]{SalomonFerrer2013}%
  \BibitemOpen
  \bibfield  {author} {\bibinfo {author} {\bibnamefont {Salomon-Ferrer},
  \bibfnamefont {R.}}, \bibinfo {author} {\bibnamefont {Case}, \bibfnamefont
  {D.~A.}}, \ and\ \bibinfo {author} {\bibnamefont {Walker}, \bibfnamefont
  {R.~C.}},\ }\href@noop {} {\bibfield  {journal} {\bibinfo  {journal} {Wiley
  Interdisciplinary Reviews: Computational Molecular Science}\ }\textbf
  {\bibinfo {volume} {3}},\ \bibinfo {pages} {198} (\bibinfo {year}
  {2013})}\BibitemShut {NoStop}%
\bibitem [{\citenamefont {Sharp}\ \emph {et~al.}(1991)\citenamefont {Sharp},
  \citenamefont {Nicholls}, \citenamefont {Friedman},\ and\ \citenamefont
  {Honig}}]{Sharp1991}%
  \BibitemOpen
  \bibfield  {author} {\bibinfo {author} {\bibnamefont {Sharp}, \bibfnamefont
  {K.~A.}}, \bibinfo {author} {\bibnamefont {Nicholls}, \bibfnamefont {a.}},
  \bibinfo {author} {\bibnamefont {Friedman}, \bibfnamefont {R.}}, \ and\
  \bibinfo {author} {\bibnamefont {Honig}, \bibfnamefont {B.}},\ }\href@noop {}
  {\bibfield  {journal} {\bibinfo  {journal} {Biochemistry}\ }\textbf {\bibinfo
  {volume} {30}},\ \bibinfo {pages} {9686} (\bibinfo {year}
  {1991})}\BibitemShut {NoStop}%
\bibitem [{\citenamefont {Smith}\ \emph {et~al.}(2012)\citenamefont {Smith},
  \citenamefont {Schreck}, \citenamefont {Hashem}, \citenamefont {Soltani},
  \citenamefont {Nath}, \citenamefont {Rhoades},\ and\ \citenamefont
  {O'Hern}}]{Smith2012}%
  \BibitemOpen
  \bibfield  {author} {\bibinfo {author} {\bibnamefont {Smith}, \bibfnamefont
  {W.~W.}}, \bibinfo {author} {\bibnamefont {Schreck}, \bibfnamefont {C.~F.}},
  \bibinfo {author} {\bibnamefont {Hashem}, \bibfnamefont {N.}}, \bibinfo
  {author} {\bibnamefont {Soltani}, \bibfnamefont {S.}}, \bibinfo {author}
  {\bibnamefont {Nath}, \bibfnamefont {A.}}, \bibinfo {author} {\bibnamefont
  {Rhoades}, \bibfnamefont {E.}}, \ and\ \bibinfo {author} {\bibnamefont
  {O'Hern}, \bibfnamefont {C.~S.}},\ }\href {\doibase
  10.1103/PhysRevE.86.041910} {\bibfield  {journal} {\bibinfo  {journal}
  {Physical Review E}\ }\textbf {\bibinfo {volume} {86}},\ \bibinfo {pages}
  {041910} (\bibinfo {year} {2012})}\BibitemShut {NoStop}%
\bibitem [{\citenamefont {Sugase}, \citenamefont {Dyson},\ and\ \citenamefont
  {Wright}(2007)}]{Sugase2007}%
  \BibitemOpen
  \bibfield  {author} {\bibinfo {author} {\bibnamefont {Sugase}, \bibfnamefont
  {K.}}, \bibinfo {author} {\bibnamefont {Dyson}, \bibfnamefont {H.~J.}}, \
  and\ \bibinfo {author} {\bibnamefont {Wright}, \bibfnamefont {P.~E.}},\
  }\href {\doibase 10.1038/nature05858} {\bibfield  {journal} {\bibinfo
  {journal} {Nature}\ }\textbf {\bibinfo {volume} {447}},\ \bibinfo {pages}
  {1021} (\bibinfo {year} {2007})}\BibitemShut {NoStop}%
\bibitem [{\citenamefont {Szil\'{a}gyi}, \citenamefont {Gy\"{o}rffy},\ and\
  \citenamefont {Z\'{a}vodszky}(2008)}]{Szilagyi2008}%
  \BibitemOpen
  \bibfield  {author} {\bibinfo {author} {\bibnamefont {Szil\'{a}gyi},
  \bibfnamefont {A.}}, \bibinfo {author} {\bibnamefont {Gy\"{o}rffy},
  \bibfnamefont {D.}}, \ and\ \bibinfo {author} {\bibnamefont {Z\'{a}vodszky},
  \bibfnamefont {P.}},\ }\href@noop {} {\bibfield  {journal} {\bibinfo
  {journal} {Biophysical journal}\ }\textbf {\bibinfo {volume} {95}},\ \bibinfo
  {pages} {1612} (\bibinfo {year} {2008})}\BibitemShut {NoStop}%
\bibitem [{\citenamefont {Tashiro}\ \emph {et~al.}(2008)\citenamefont
  {Tashiro}, \citenamefont {Kojima}, \citenamefont {Kihara}, \citenamefont
  {Kasai}, \citenamefont {Kamiyoshihara}, \citenamefont {Uéda},\ and\
  \citenamefont {Shimotakahara}}]{tashiro2008}%
  \BibitemOpen
  \bibfield  {author} {\bibinfo {author} {\bibnamefont {Tashiro}, \bibfnamefont
  {M.}}, \bibinfo {author} {\bibnamefont {Kojima}, \bibfnamefont {M.}},
  \bibinfo {author} {\bibnamefont {Kihara}, \bibfnamefont {H.}}, \bibinfo
  {author} {\bibnamefont {Kasai}, \bibfnamefont {K.}}, \bibinfo {author}
  {\bibnamefont {Kamiyoshihara}, \bibfnamefont {T.}}, \bibinfo {author}
  {\bibnamefont {Uéda}, \bibfnamefont {K.}}, \ and\ \bibinfo {author}
  {\bibnamefont {Shimotakahara}, \bibfnamefont {S.}},\ }\href {\doibase
  10.1016/j.bbrc.2008.02.127} {\bibfield  {journal} {\bibinfo  {journal}
  {Biochemical and Biophysical Research Communications}\ }\textbf {\bibinfo
  {volume} {369}},\ \bibinfo {pages} {910} (\bibinfo {year}
  {2008})}\BibitemShut {NoStop}%
\bibitem [{\citenamefont {Trexler}\ and\ \citenamefont
  {Rhoades}(2013)}]{trexler2013}%
  \BibitemOpen
  \bibfield  {author} {\bibinfo {author} {\bibnamefont {Trexler}, \bibfnamefont
  {A.}}\ and\ \bibinfo {author} {\bibnamefont {Rhoades}, \bibfnamefont {E.}},\
  }\href@noop {} {\bibfield  {journal} {\bibinfo  {journal} {Mol. Neurobiol.}\
  }\textbf {\bibinfo {volume} {47}},\ \bibinfo {pages} {622} (\bibinfo {year}
  {2013})}\BibitemShut {NoStop}%
\bibitem [{\citenamefont {Trexler}\ and\ \citenamefont
  {Rhoades}(2010)}]{Trexler2010}%
  \BibitemOpen
  \bibfield  {author} {\bibinfo {author} {\bibnamefont {Trexler}, \bibfnamefont
  {A.~J.}}\ and\ \bibinfo {author} {\bibnamefont {Rhoades}, \bibfnamefont
  {E.}},\ }\href {\doibase 10.1016/j.bpj.2010.08.056} {\bibfield  {journal}
  {\bibinfo  {journal} {Biophysical journal}\ }\textbf {\bibinfo {volume}
  {99}},\ \bibinfo {pages} {3048} (\bibinfo {year} {2010})}\BibitemShut
  {NoStop}%
\bibitem [{\citenamefont {U\'{e}da}\ \emph {et~al.}(1993)\citenamefont
  {U\'{e}da}, \citenamefont {Fukushima}, \citenamefont {Masliah}, \citenamefont
  {Xia}, \citenamefont {Iwai}, \citenamefont {Yoshimoto}, \citenamefont
  {Otero}, \citenamefont {Kondo}, \citenamefont {Ihara},\ and\ \citenamefont
  {Saitoh}}]{Ueda1993}%
  \BibitemOpen
  \bibfield  {author} {\bibinfo {author} {\bibnamefont {U\'{e}da},
  \bibfnamefont {K.}}, \bibinfo {author} {\bibnamefont {Fukushima},
  \bibfnamefont {H.}}, \bibinfo {author} {\bibnamefont {Masliah}, \bibfnamefont
  {E.}}, \bibinfo {author} {\bibnamefont {Xia}, \bibfnamefont {Y.}}, \bibinfo
  {author} {\bibnamefont {Iwai}, \bibfnamefont {a.}}, \bibinfo {author}
  {\bibnamefont {Yoshimoto}, \bibfnamefont {M.}}, \bibinfo {author}
  {\bibnamefont {Otero}, \bibfnamefont {D.~a.}}, \bibinfo {author}
  {\bibnamefont {Kondo}, \bibfnamefont {J.}}, \bibinfo {author} {\bibnamefont
  {Ihara}, \bibfnamefont {Y.}}, \ and\ \bibinfo {author} {\bibnamefont
  {Saitoh}, \bibfnamefont {T.}},\ }\href@noop {} {\bibfield  {journal}
  {\bibinfo  {journal} {Proceedings of the National Academy of Sciences of the
  United States of America}\ }\textbf {\bibinfo {volume} {90}},\ \bibinfo
  {pages} {11282} (\bibinfo {year} {1993})}\BibitemShut {NoStop}%
\bibitem [{\citenamefont {Uversky}, \citenamefont {Gillespie},\ and\
  \citenamefont {Fink}(2000)}]{Uversky2000}%
  \BibitemOpen
  \bibfield  {author} {\bibinfo {author} {\bibnamefont {Uversky}, \bibfnamefont
  {V.~N.}}, \bibinfo {author} {\bibnamefont {Gillespie}, \bibfnamefont
  {J.~R.}}, \ and\ \bibinfo {author} {\bibnamefont {Fink}, \bibfnamefont
  {A.~L.}},\ }\href@noop {} {\bibfield  {journal} {\bibinfo  {journal}
  {Proteins}\ }\textbf {\bibinfo {volume} {41}},\ \bibinfo {pages} {415}
  (\bibinfo {year} {2000})}\BibitemShut {NoStop}%
\bibitem [{\citenamefont {Uversky}, \citenamefont {Li},\ and\ \citenamefont
  {Fink}(2001)}]{uversky2001}%
  \BibitemOpen
  \bibfield  {author} {\bibinfo {author} {\bibnamefont {Uversky}, \bibfnamefont
  {V.~N.}}, \bibinfo {author} {\bibnamefont {Li}, \bibfnamefont {J.}}, \ and\
  \bibinfo {author} {\bibnamefont {Fink}, \bibfnamefont {A.~L.}},\ }\href
  {\doibase 10.1016/S0014-5793(01)03121-0} {\bibfield  {journal} {\bibinfo
  {journal} {{FEBS} Letters}\ }\textbf {\bibinfo {volume} {509}},\ \bibinfo
  {pages} {31} (\bibinfo {year} {2001})}\BibitemShut {NoStop}%
\bibitem [{\citenamefont {Uversky}\ \emph {et~al.}(2002)\citenamefont
  {Uversky}, \citenamefont {Li}, \citenamefont {Souillac}, \citenamefont
  {Millett}, \citenamefont {Doniach}, \citenamefont {Jakes}, \citenamefont
  {Goedert},\ and\ \citenamefont {Fink}}]{Uversky2002b}%
  \BibitemOpen
  \bibfield  {author} {\bibinfo {author} {\bibnamefont {Uversky}, \bibfnamefont
  {V.~N.}}, \bibinfo {author} {\bibnamefont {Li}, \bibfnamefont {J.}}, \bibinfo
  {author} {\bibnamefont {Souillac}, \bibfnamefont {P.}}, \bibinfo {author}
  {\bibnamefont {Millett}, \bibfnamefont {I.~S.}}, \bibinfo {author}
  {\bibnamefont {Doniach}, \bibfnamefont {S.}}, \bibinfo {author} {\bibnamefont
  {Jakes}, \bibfnamefont {R.}}, \bibinfo {author} {\bibnamefont {Goedert},
  \bibfnamefont {M.}}, \ and\ \bibinfo {author} {\bibnamefont {Fink},
  \bibfnamefont {A.~L.}},\ }\href {\doibase 10.1074/jbc.M109541200} {\bibfield
  {journal} {\bibinfo  {journal} {The Journal of biological chemistry}\
  }\textbf {\bibinfo {volume} {277}},\ \bibinfo {pages} {11970} (\bibinfo
  {year} {2002})}\BibitemShut {NoStop}%
\bibitem [{\citenamefont {Uversky}\ \emph {et~al.}(2005)\citenamefont
  {Uversky}, \citenamefont {Yamin}, \citenamefont {Munishkina}, \citenamefont
  {Karymov}, \citenamefont {Millett}, \citenamefont {Doniach}, \citenamefont
  {Lyubchenko},\ and\ \citenamefont {Fink}}]{uversky2005}%
  \BibitemOpen
  \bibfield  {author} {\bibinfo {author} {\bibnamefont {Uversky}, \bibfnamefont
  {V.~N.}}, \bibinfo {author} {\bibnamefont {Yamin}, \bibfnamefont {G.}},
  \bibinfo {author} {\bibnamefont {Munishkina}, \bibfnamefont {L.~A.}},
  \bibinfo {author} {\bibnamefont {Karymov}, \bibfnamefont {M.~A.}}, \bibinfo
  {author} {\bibnamefont {Millett}, \bibfnamefont {I.~S.}}, \bibinfo {author}
  {\bibnamefont {Doniach}, \bibfnamefont {S.}}, \bibinfo {author} {\bibnamefont
  {Lyubchenko}, \bibfnamefont {Y.~L.}}, \ and\ \bibinfo {author} {\bibnamefont
  {Fink}, \bibfnamefont {A.~L.}},\ }\href {\doibase
  10.1016/j.molbrainres.2004.11.014} {\bibfield  {journal} {\bibinfo  {journal}
  {Molecular Brain Research}\ }\textbf {\bibinfo {volume} {134}},\ \bibinfo
  {pages} {84} (\bibinfo {year} {2005})}\BibitemShut {NoStop}%
\bibitem [{\citenamefont {Vucetic}\ \emph {et~al.}(2003)\citenamefont
  {Vucetic}, \citenamefont {Brown}, \citenamefont {Dunker},\ and\ \citenamefont
  {Obradovic}}]{Vucetic2003}%
  \BibitemOpen
  \bibfield  {author} {\bibinfo {author} {\bibnamefont {Vucetic}, \bibfnamefont
  {S.}}, \bibinfo {author} {\bibnamefont {Brown}, \bibfnamefont {C.~J.}},
  \bibinfo {author} {\bibnamefont {Dunker}, \bibfnamefont {A.~K.}}, \ and\
  \bibinfo {author} {\bibnamefont {Obradovic}, \bibfnamefont {Z.}},\
  }\href@noop {} {\bibfield  {journal} {\bibinfo  {journal} {Proteins}\
  }\textbf {\bibinfo {volume} {52}},\ \bibinfo {pages} {573} (\bibinfo {year}
  {2003})}\BibitemShut {NoStop}%
\bibitem [{\citenamefont {Wang}\ and\ \citenamefont
  {Dunbrack}(2003)}]{Wang2003}%
  \BibitemOpen
  \bibfield  {author} {\bibinfo {author} {\bibnamefont {Wang}, \bibfnamefont
  {G.}}\ and\ \bibinfo {author} {\bibnamefont {Dunbrack}, \bibfnamefont
  {R.~L.}},\ }\href {\doibase 10.1093/bioinformatics/btg224} {\bibfield
  {journal} {\bibinfo  {journal} {Bioinformatics}\ }\textbf {\bibinfo {volume}
  {19}},\ \bibinfo {pages} {1589} (\bibinfo {year} {2003})},\ \bibinfo {note}
  {pMID: 12912846}\BibitemShut {NoStop}%
\bibitem [{\citenamefont {White}\ and\ \citenamefont
  {Wimley}(1999)}]{White1999}%
  \BibitemOpen
  \bibfield  {author} {\bibinfo {author} {\bibnamefont {White}, \bibfnamefont
  {S.~H.}}\ and\ \bibinfo {author} {\bibnamefont {Wimley}, \bibfnamefont
  {W.~C.}},\ }\href {\doibase 10.1146/annurev.biophys.28.1.319} {\bibfield
  {journal} {\bibinfo  {journal} {Annual review of biophysics and biomolecular
  structure}\ }\textbf {\bibinfo {volume} {28}},\ \bibinfo {pages} {319}
  (\bibinfo {year} {1999})}\BibitemShut {NoStop}%
\bibitem [{\citenamefont {Wimley}\ and\ \citenamefont
  {White}(1996)}]{Wimley1996}%
  \BibitemOpen
  \bibfield  {author} {\bibinfo {author} {\bibnamefont {Wimley}, \bibfnamefont
  {W.~C.}}\ and\ \bibinfo {author} {\bibnamefont {White}, \bibfnamefont
  {S.~H.}},\ }\href@noop {} {\bibfield  {journal} {\bibinfo  {journal} {Nature
  Structural Biology}\ }\textbf {\bibinfo {volume} {3}},\ \bibinfo {pages}
  {842} (\bibinfo {year} {1996})}\BibitemShut {NoStop}%
\end{thebibliography}%

\end{multicols}

\end{document}